\documentclass{article}

\usepackage{arxiv}

\usepackage[utf8]{inputenc} 
\usepackage[T1]{fontenc}    
\usepackage{hyperref}       
\usepackage{url}            
\usepackage{booktabs}       
\usepackage{amsfonts}       
\usepackage{nicefrac}       
\usepackage{microtype}      
\usepackage{lipsum}
\usepackage{graphicx}
\usepackage{amsmath}
\usepackage{makecell}
\graphicspath{ {./images/} }

\title{A Hybrid CNN-Cheby-KAN Framework for Efficient Prediction of Two-Dimensional Airfoil Pressure Distribution}

\author{
 Yaohong Chen \\
  AVIC Aerodynamics Research Institute\\
  Harbin, China 150001 \\
  \texttt{marlboroiron@163.com}
  \And
 Luchi Zhang \\
  China Gas Turbine Establishment\\
  Chengdu, China 610599 \\
  \And
 Yiju Deng \\
  Graduate School of Chinese Aeronautical Establishment\\
  Yangzhou, China 225111 \\
  \And
 Yanze Yu\\
  Graduate School of Chinese Aeronautical Establishment\\
  Yangzhou, China 225111 \\
  \And
 Xiang Li\\
  AVIC Aerodynamics Research Institute\\
  Shenyang, China 110034 \\
  \And
 Renshan Jiao\\
  AVIC Aerodynamics Research Institute\\
  Harbin, China 150001 \\
  \texttt{caria\_jiao@126.com}
}

\begin{document}
\maketitle
\begin{abstract}
The accurate prediction of airfoil pressure distribution is essential for aerodynamic performance evaluation, yet traditional methods such as computational fluid dynamics (CFD) and wind tunnel testing have certain bottlenecks. This paper proposes a hybrid deep learning model combining a Convolutional Neural Network (CNN) and a Chebyshev-enhanced Kolmogorov-Arnold Network (Cheby-KAN) for efficient and accurate prediction of the two-dimensional airfoil flow field. The CNN learns 1549 types of airfoils and encodes airfoil geometries into a compact 16-dimensional feature vector, while the Cheby-KAN models complex nonlinear mappings from flight conditions and spatial coordinates to pressure values. Experiments on multiple airfoils—including RAE2822, NACA0012, e387, and mh38—under various Reynolds numbers and angles of attack demonstrate that the proposed method achieves a mean squared error (MSE) on the order of $10^{-6}$ and a coefficient of determination ($R^2$) exceeding 0.999. The model significantly outperforms traditional Multilayer Perceptrons (MLPs) in accuracy and generalizability, with acceptable computational overhead. These results indicate that the hybrid CNN-Cheby-KAN framework offers a promising data-driven approach for rapid aerodynamic prediction.
\end{abstract}


\section{Introduction}
Aerodynamic design plays a crucial role in modern aerospace, wind energy, and automotive industries. Accurate analysis and performance evaluation of the flow field around an airfoil—such as pressure and velocity distributions—form the foundation for efficient and high-performance aerodynamic design \cite{tan2015influence} \cite{bhatnagar2019prediction}. Current mainstream methods for obtaining pressure distribution face significant bottlenecks: wind tunnel tests rely on sparse sensor arrays and are susceptible to multiple interference factors, while computational fluid dynamics (CFD) simulations are computationally expensive and time-consuming. Breakthroughs in data science and deep learning are driving the transformation of flow field modeling from physics-driven to data-driven with physical constraints approaches, and these emerging methodologies are expected to overcome the limitations of both traditional methods\cite{hey2009fourth}.

In recent years, numerous researchers have integrated machine learning theories into airfoil aerodynamic research, aiming to conduct aerodynamic optimization and shape design in a faster and more cost-effective manner\cite{Peng2022DeepNN}\cite{xiao2024learning}\cite{lou2023aerodynamic}. For instance, Obiols-Sales et al.\cite{obiols2020cfdnet} developed the CFDnet framework, which incorporates a convolutional neural network (CNN) to efficiently solve the Reynolds-averaged Navier-Stokes equations, achieving a speedup of two orders of magnitude in flow field prediction. Mi Baigang et al.\cite{mi2025intelligent} proposed an improved U-Net-LSTM hybrid neural network that enhances low-dimensional geometric and operational condition information through Signed Distance Function (SDF) and composite images, enabling efficient modeling of both steady and unsteady flow fields. Their experimental results demonstrated prediction errors of less than 1.98\% and 2.56\% under steady and unsteady conditions, respectively. Nils Thuerey et al.\cite{thuerey2020deep} systematically evaluated numerous trained neural networks based on a modified U-Net architecture for predicting pressure and velocity fields. Their study elucidated the influence of training dataset size and model parameters on predictive accuracy, achieving mean relative errors below 3\% across various unseen airfoil configurations. Tompson et al.\cite{tompson2017accelerating} introduced an innovative approach using convolutional neural networks that combine local convolution operations with global downsampling and upsampling structures to capture long-range physical phenomena (such as pressure gradients) and local flow details. This method significantly improves the modeling of long-range effects compared to traditional U-Net models, thereby better capturing long-range pressure propagation effects, a key limitation of local convolution operations in standard U-Nets. Additionally, many researchers\cite{wang2023airfoil}\cite{zhang2025advances}\cite{esfahanian2024aerodynamic} have developed machine learning-based surrogate models for pressure distribution prediction and applied them to establish efficient aerodynamic optimization frameworks.

Despite these advances, deep learning architectures such as CNNs and multi-layer perceptron (MLPs) still face several challenges. First, there is still a discrepancy between the predicted results of the flow field and the CFD simulation results. For example, in the boundary layer region, even a slight error in the predicted field can lead to significant errors in gradient calculation\cite{sekar2019fast}. Second, due to inherent limitations in the architecture of CNNs and other models, the computational cost and training time for models trained on large datasets increase significantly. For sparse flow field data, Zuo et al.\cite{zuo2022fast} used the Multi-Hierarchical Perceptron(MHP) to decouple the tasks, which improved the model's accuracy and achieved good results with small samples. However, the parameter efficiency of their architecture was not significantly improved, motivating our exploration of more parameter-efficient alternatives like Cheby-KAN.

To address these issues, this paper proposes a novel flow field prediction method for two-dimensional airfoils based on a combination of CNN, Chebyshev polynomials, and the Kolmogorov-Arnold Network(KAN). This approach eliminates the reliance of traditional parameterization techniques such as Class function/Shape function Transformation (CST) and Non-Uniform Rational B-Splines (NURBS) on manual parameter tuning. By automatically extracting a low-dimensional feature vector of the two-dimensional airfoil, this method achieves an automated process that converts original geometric data into a compact feature representation. The innovative KAN architecture serves as the core of the prediction model, significantly enhancing interpretability and reducing the number of parameters\cite{liu2024kan}. Moreover, Chebyshev polynomials are used instead of the original spline basis functions to further accelerate convergence and improve fitting performance. Experimental results demonstrate that the proposed model exhibits superior approximation capability compared to MLP-based methods, particularly within the boundary layer and around the trailing edge of the airfoil.
\section{Methodology}
\subsection{Deep learning method}
\subsubsection{Convolutional Neural Network}
CNNs represent a class of deep learning architectures specifically tailored for processing grid-structured data, such as images or flow field meshes in fluid dynamics. Their design leverages key principles such as local connectivity, weight sharing, and spatial pooling, which collectively enable the efficient extraction of hierarchical features from input data. Unlike traditional fully connected neural networks, CNNs utilize learnable filters (or kernels), which perform localized weighted summations as they convolve across the input. This approach significantly reduces the number of parameters while effectively capturing translation-invariant local patterns. These characteristics provide CNNs with considerable advantages in modeling low-dimensional embeddings and characterizing flow field features in fluid dynamics.

\begin{figure} 
    \centering
    \includegraphics[width=8cm]{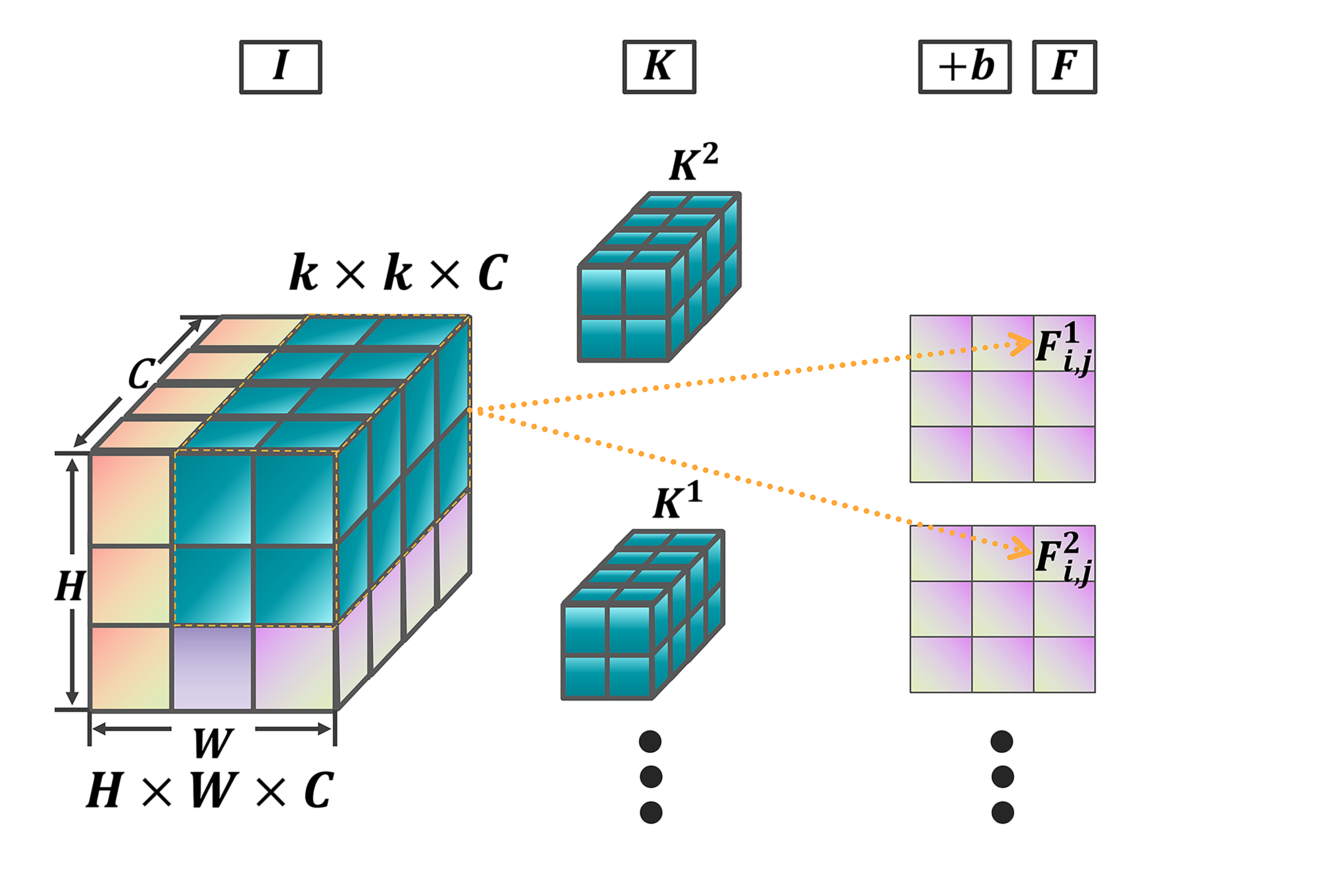}
    \caption{Convolution operations under multiple convolution kernels}
    \label{fig:Fig1}
\end{figure}

A typical CNN architecture comprises convolutional layers, pooling layers, and activation functions. The convolutional layer applies these filters through convolution operations to generate local feature maps.\cite{lecun2002gradient} The value at position $(i,j)$ in the output feature map $F$ for a 2D input image $I \in R^{H \times W \times C}$ (of height $H$, width $W$, channels $C$) using a convolution kernel $K \in R^{k \times k \times c}$ is computed as:
\begin{equation}
  F(i,j) = \sum_{c=0}^{C} \sum_{u=0}^{k-1} \sum_{v=0}^{k-1} K(u,v,c) \cdot I(i+u,j+v,c) + b_c
\end{equation}
In this formulation, $k$ denotes the convolution kernel size and $b$ is a bias term. As shown in Figure\ref{fig:Fig1}, multiple kernels can be applied in parallel to produce multi-channel feature maps, enabling the simultaneous extraction of diverse features. 

\begin{figure} 
    \centering
    \includegraphics[width=8cm]{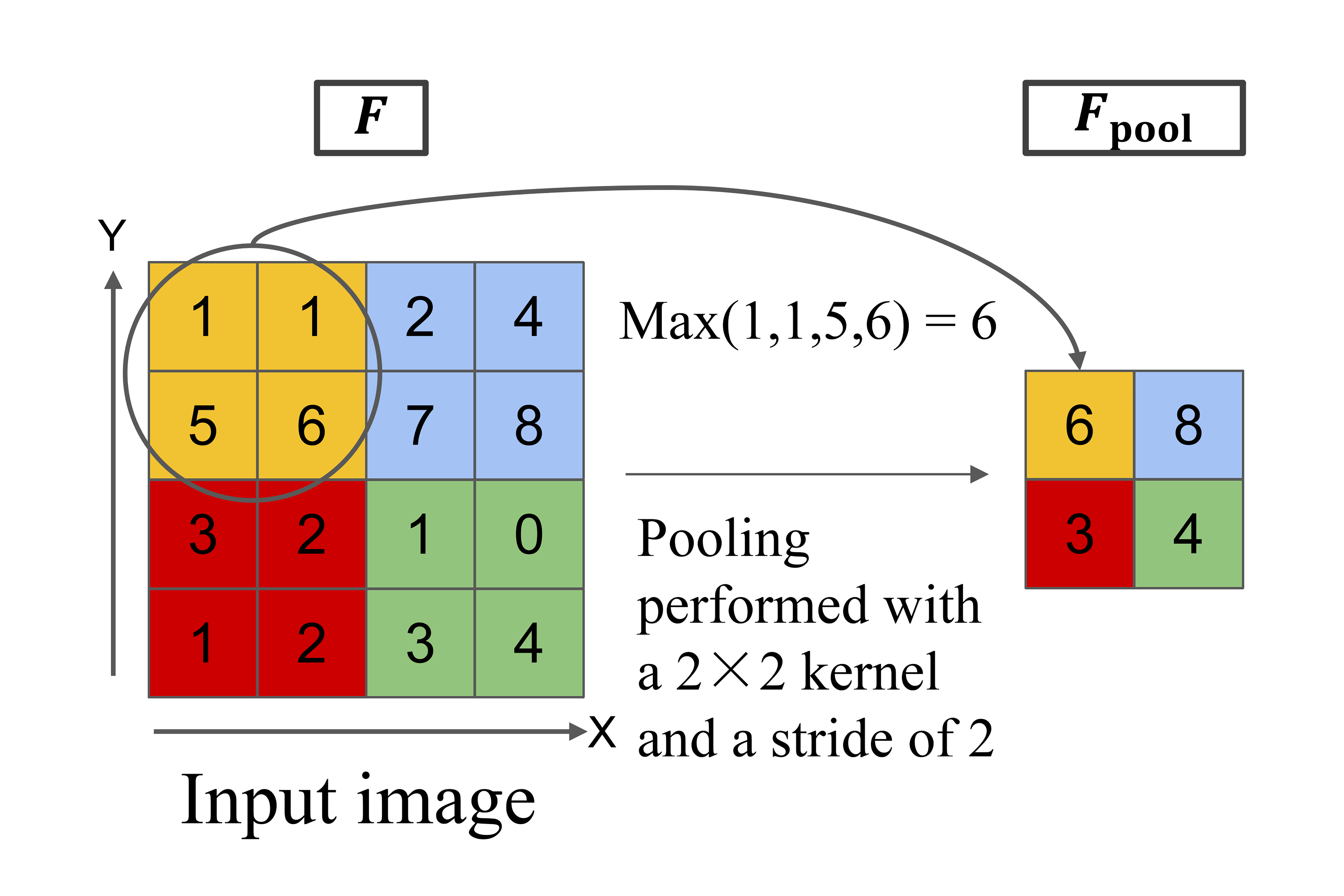}
    \caption{Schematic diagram of maximum pooling principle}
    \label{fig:Fig2}
\end{figure}

Subsequent to convolution, pooling operations are employed to reduce the feature map dimensionality and enhance the model's invariance to geometric distortions by capturing dominant features within local regions. Common approaches include max pooling and average pooling. The max pooling operation, for example, can be formulated as:
\begin{equation}
  F_{pool}(i,j,d) = \max_{u,v \in N(i,j)} F(i+u,j+v,d)
\end{equation}
In this formulation, $N(i,j)$ denotes local region (e.g., a $k \times k$ window) centered at location $(i,j)$ in the input feature map $F$, shown in Figure \ref{fig:Fig2}. This operation selects the maximum value within the specified region, effectively downsampling the feature map while preserving the most prominent features. By stacking multiple convolution and pooling layers, CNNs can progressively extract increasingly abstract and complex features from the input data, ultimately enabling the effective representation and prediction of flow field patterns.

To model complex nonlinear relationships, CNNs incorporate nonlinear activation functions, such as ReLU and Leaky ReLU. The ReLU function is defined as:
\begin{equation}
  \sigma = \max(0,x)
\end{equation}
This sparse activation mechanism helps mitigate the vanishing gradient problem and promotes computational efficiency\cite{khalid2020empirical}.

\begin{figure} 
    \centering
    \includegraphics[width=14cm]{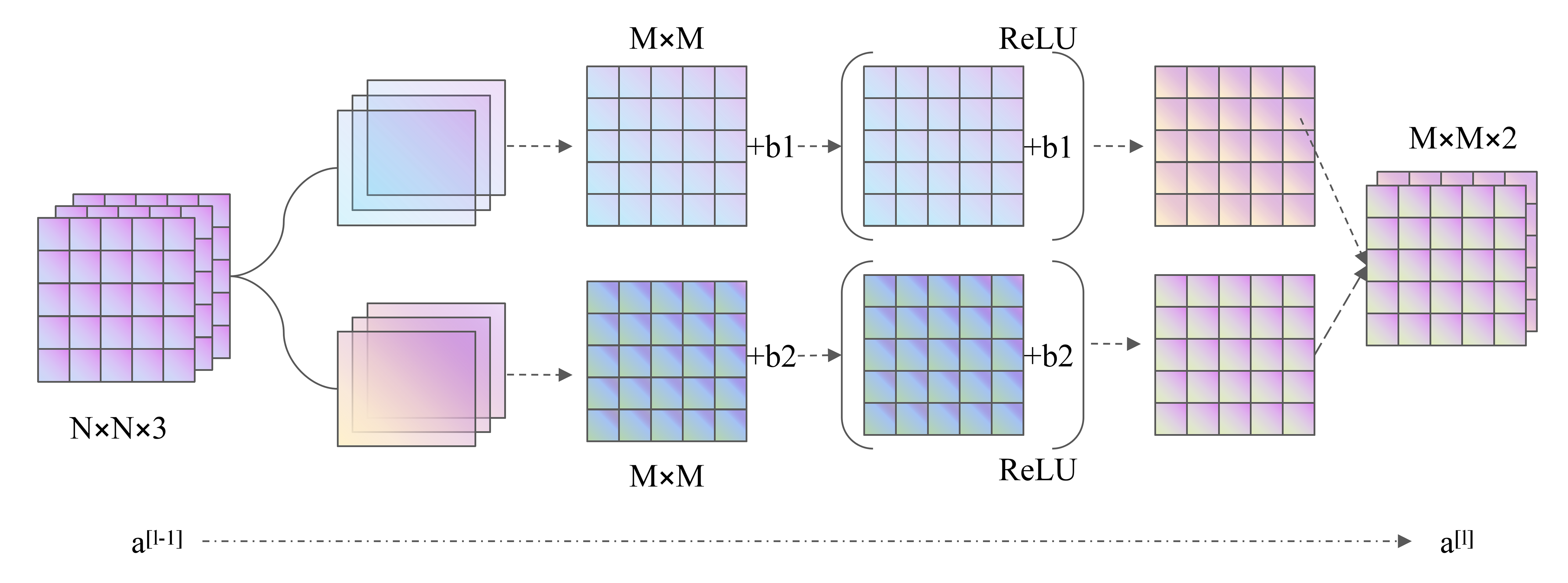}
    \caption{The complete structure of convolutional neural networks}
    \label{fig:Fig3}
\end{figure}

Finally, by appropriately combining and stacking convolutional, pooling, and activation layers as required, a complete convolutional neural network (CNN) is constructed (Figure \ref{fig:Fig3}). In this study, we designed a CNN augmented with fully-connected layers to automatically extract discriminative feature representations from airfoil data. The detailed architecture and hyperparameter configurations of the network are provided in Section \ref{sec:sec2.2.1}.

\subsubsection{Kolmogorov-Arnold Networks}
Kolmogorov-Arnold Networks (KANs) are a novel neural network architecture inspired by the Kolmogorov-Arnold representation theorem. This theorem states that any multivariate continuous function $f:[0,1]^n \to \mathbb{R}$ defined on a bounded domain can be represented as a finite composition of continuous univariate functions and addition operations. Specifically, the theorem guarantees that $f$ can be written in the following form:
\begin{equation}
  f(x) = f(x_1,x_2,x_3,\dots,x_n) = \sum_{q=1}^{2n+1} \phi_q(\sum_{p=1}^{n}\phi_{q,p}(x_p))
\end{equation}
where each $\phi_{q,p}:[0,1] \to \mathbb{R}$ and $\phi_q:\mathbb{R} \to \mathbb{R}$ is a continuous univariate function. This result demonstrates that the structure of any multivariate continuous function can be fundamentally decomposed into additions and compositions of univariate functions.

\begin{figure} 
    \centering
    \includegraphics[width=8cm]{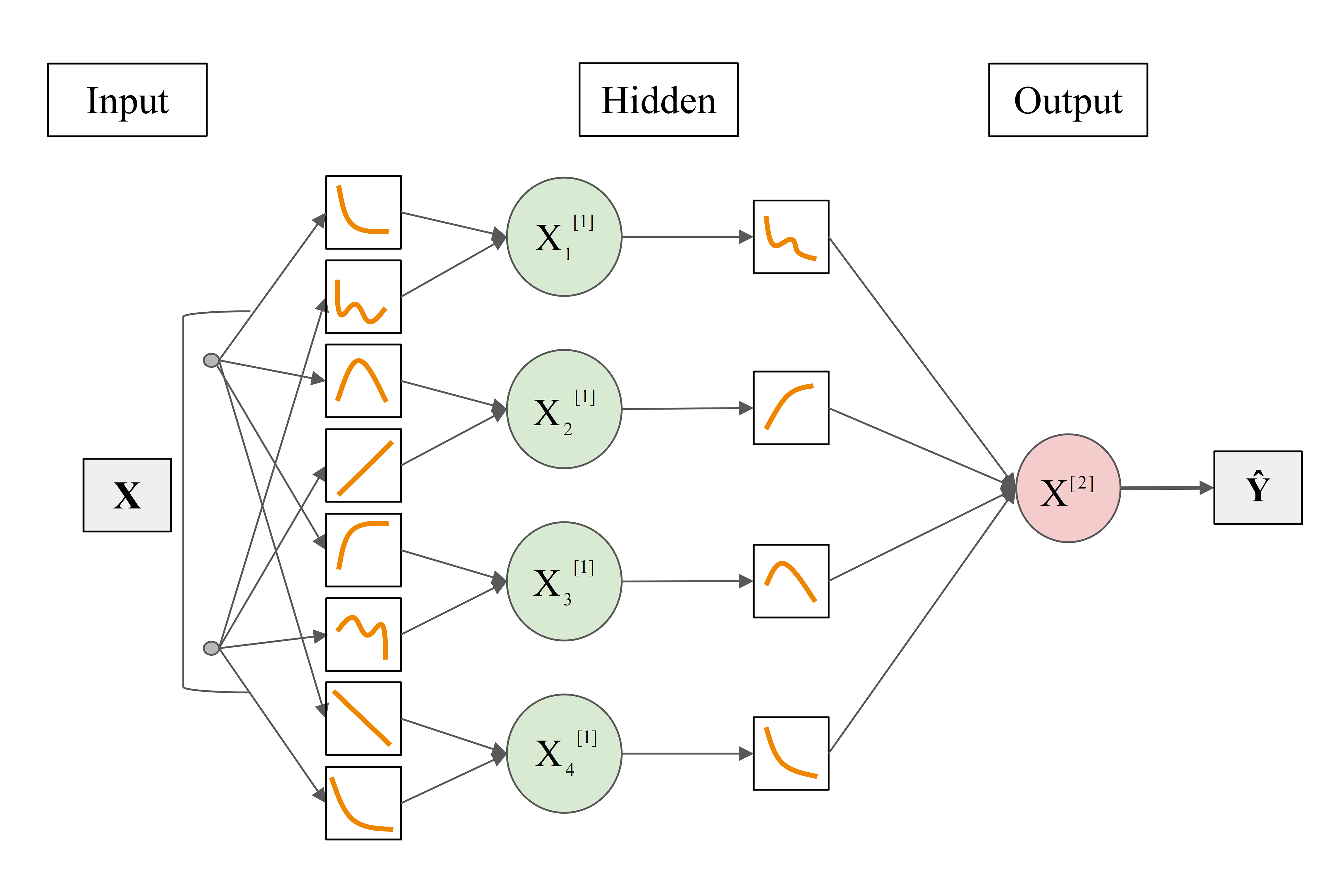}
    \caption{Schematic diagram of Kolmogorov Arnold networks structure}
    \label{fig:Fig4}
\end{figure}

Inspired by this theorem, KANs are designed as a class of neural networks that explicitly parameterize the functional decomposition described in the theorem (Figure \ref{fig:Fig4}). In contrast to Multi-Layer Perceptrons (MLPs), which apply fixed activation functions at the nodes, KANs employ learnable univariate activation functions on the edges. More specifically, each connection in a KAN substitutes a conventional weight with a univariate function, commonly parameterized using a B-spline curve with trainable coefficients. The nodes themselves only perform summation operations and do not incorporate any nonlinear activation.

Mathematically, each learnable activation function $\phi(x)$ is composed of a combination of basis functions and differentiable splines:
\begin{equation}
  \phi(x) = w_b \cdot  b(x) + w_s \cdot spline(x)
\end{equation}
where the basis function $b(x)$ is chosen as the SiLU (Sigmoid Linear Unit) function:
\begin{equation}
  b(x) = SiLu(x) = \frac{x}{1+e^{-x}}
\end{equation}
and $spline(x)$ is a linear combination of B-spline basis functions:
\begin{equation}
  spline(x) = \sum_{i}c_iB_i(x)
\end{equation}
where $B_i(x)$ is the $i$-th B-spline basis function, and $c_i$ are the trainable coefficients. The SiLU function is chosen due to its smoothness and the ability to approximate any continuous function, while the B-spline basis functions provide a flexible and smooth representation of the activation functions.

This combination of a basis function and a spline provides a flexible nonlinear transformation: the basis function offers a smooth initial mapping, while the spline component allows fine-grained local adjustments. Consequently, this enhances the model's expressive power and enables it to capture complex, non-uniform patterns in the data.

Relevant research\cite{hollosi2025driver}\cite{urbanczyk2025kolmogorov}\cite{wang2025intrusion} have demonstrated that KANs achieve higher accuracy than traditional MLPs in various tasks, often with fewer parameters. However, this gain in accuracy and parameter efficiency may come at the cost of increased computational and memory requirements, making current implementations of KANs generally slower and more resource-intensive than MLPs.

Beyond their advantages in accuracy, KANs also offer improved interpretability. The use of univariate edge functions makes it easier to visualize and understand the network's behavior, and techniques such as pruning and symbolic regularization can further simplify the model. These features make KANs particularly promising for scientific applications where model transparency and the discovery of symbolic relationships are essential.
\subsubsection{Chebyshev polynomials}
Chebyshev polynomials are a class of orthogonal polynomials widely used in mathematics and engineering. There are two main types: Chebyshev polynomials of the first kind, denoted as $T_n(x)$, and those of the second kind, denoted as $U_n(x)$. This study focuses primarily on Chebyshev polynomials of the first kind.

The Chebyshev polynomials of the first kind constitute a sequence of orthogonal polynomials with respect to the weight function $\sqrt{(1-x^2)}$ over the interval $[-1,1]$. They can be defined by the recurrence relation:
\begin{equation}
  \begin{split}
    T_0(x) &= 1 \\ 
    T_1(x) &= x \\ 
    T_{n}(x) &= 2x \cdot T_{n-1}(x) - T_{n-2}(x), n \geq 2
  \end{split}
\end{equation}
Chebyshev polynomials possess a number of important mathematical properties, including orthogonality, recursive computability, and minimax optimality on the interval [-1,1]. These characteristics make them particularly suitable for numerical approximation and feature transformation in machine learning models.

\paragraph{Orthogonality.}
Chebyshev polynomials satisfy the following orthogonality condition on the interval $[-1,1]$ with respect to the weight function $\omega(x) = 1/\sqrt{(1-x^2)}$. For non-negative integers $m$ and $n$ with $m \neq n$:
\begin{equation}
  \int_{-1}^{1} \frac{T_m(x)T_n(x)}{\sqrt{(1-x^2)}}dx = 0
\end{equation}
\paragraph{Extremal Properties.}
On the interval $[-1,1]$, the Chebyshev polynomials $T_n(x)$ attain their extreme values of $+1$ and $-1$ at the Chebyshev nodes $x_k = \cos (k\pi/n)$ for $k = 0,1,\ldots ,n$.This bounded oscillation makes them highly valuable in approximation theory, as they help mitigate Runge's phenomenon—the undesirable oscillation of high-degree polynomial interpolants near the endpoints of an interval. This characteristic is particularly beneficial in machine learning contexts, where employing Chebyshev polynomials can improve a model's ability to capture complex nonlinear relationships, thereby enhancing performance in tasks such as airfoil pressure prediction.
\paragraph{Minimax property.}
Among all monic polynomials (polynomials with leading coefficient 1) of degree $n$, the scaled Chebyshev polynomial $\widetilde{T}_n(x) = 1/2^{n-1} \cdot T_n(x) $ has the smallest maximum absolute value on $[-1,1]$. This minimax property ensures that Chebyshev polynomials provide nearly optimal uniform approximations to continuous functions, enabling high approximation accuracy with fewer terms—a feature especially advantageous in numerical computations and function approximation tasks.

Owing to these properties, Chebyshev polynomials offer several advantages as activation functions in neural networks. First, their orthogonality facilitates more efficient separation and combination of features within the network's latent space, thereby enhancing the model's representational capacity. Second, the minimax property ensures that approximation errors are uniformly minimized across the entire input interval, effectively mitigating the Runge phenomenon often encountered in high-degree polynomial interpolation. This contributes to improved stability during network training. Furthermore, due to their ability to achieve high-precision approximations of complex functions even at low degrees, Chebyshev polynomials can help reduce the number of network parameters while maintaining—or even improving—predictive performance. As a result, they enable the construction of more efficient and compact neural architectures.
\subsubsection{Chebyshev polynomial enhanced feedforward neural network}
In this study, we integrate the advantageous properties of Chebyshev polynomials with the Kolmogorov–Arnold Network (KAN) framework to construct a Chebyshev-polynomial-based KAN (Cheby-KAN) for predicting the pressure field around airfoils. This architecture retains the strengths of KANs in modeling complex nonlinear relationships while fully leveraging the orthogonality and minimax properties of Chebyshev polynomials. As a result, it achieves high precision in high-order polynomial interpolation with enhanced numerical stability.

The proposed Cheby-KAN is built by stacking multiple Cheby-KAN layers. For a Cheby-KAN layer, the input vector $x_i$ and the activation value $y_j$ of the j-th output neuron can be expressed using the following mathematical formula:
\begin{equation}
  y_j = y^{BasicL}(x_i) + y^{ChebyshevT}(x_i) + b_j
\end{equation}
Where $y^{BasicL}(x)$ is a linear basis transformation, and $y^{ChebyshevT}(x)$ denotes an element-wise Chebyshev polynomial transformation.

Specifically, the computation within each layer consists of the following components:

A basic linear transformation, analogous to the linear operation in a standard KAN layer, but first perform element by element nonlinear transformations on the input, is defined as:
\begin{equation}
  y_{j}^{BasicL}(x_i) = \sum_{i=1}^{n} W_{j,i}^{BasicL} \cdot \sigma(x_i)
\end{equation}
Where $W_{j,i}^{BasicL}$ is a basic weight matrix, and $\sigma(x)$ denotes a nonlinear activation function (e.g., ReLU or a Sigmoid function).

A Chebyshev polynomial transformation is subsequently applied:
\begin{equation}
  y_{j}^{ChebyshevT}(x_i) = \sum_{i=1}^{n} \sum_{k=0}^{K} c_{j,i,k} \cdot T_{k}(\hat{x}_i)
\end{equation}
Where $c_{j,i,k}$ denotes the Chebyshev coefficient. This is a three-dimensional tensor that determines how the j-th output neuron "perceives" the k-th order Chebyshev polynomial of the i-th input feature. $\hat{x}_i$ represents the input element $x_i$ scaled to the interval $[-1,1]$ using the hyperbolic tangent function, $T_k(x)$ represents the Chebyshev polynomial of degree $k$, and $K$ is the highest polynomial order used.

Finally, the output of the layer is given by:
\begin{equation}
  y_{j} = \sum_{i=1}^{n} W_{j,i}^{BasicL} \cdot \sigma(x_i) + \sum_{i=1}^{n} \sum_{k=0}^{K} c_{j,i,k} \cdot T_{k}(\tanh (x_i)) + b
\end{equation}
Where $b$ is a bias vector.

The Chebyshev polynomial transformation serves as the core component of the Cheby-KAN architecture. It is implemented through the following steps:
\begin{enumerate}
    \item Input Normalization: The input $z$ is first normalized to the interval $[-1,1]$ to align with the natural domain of Chebyshev polynomials.
    \item Chebyshev Polynomial Evaluation: Each Chebyshev polynomial $T_k(\hat{x}_i)$ can be efficiently computed using the trigonometric definition:
    \begin{equation}
      \begin{split}
        T_k(\hat{x}_i) &= \cos(k \cdot \arccos(\theta_i))\\
        \theta_i &= \arccos(\hat{x}_i)
      \end{split}
    \end{equation}
    This formulation allows stable and efficient computation of high-order polynomials.
    \item Polynomial Combination: The transformation maps the normalized input to a high-dimensional feature space through a learnable linear combination of Chebyshev polynomial terms:
    \begin{equation}
      u = \sum_{k=0}^{K} c_{k} \cdot T_k(z)
    \end{equation}
    This combination enables the network to capture sophisticated nonlinear relationships in the data.
\end{enumerate}

By leveraging the orthogonality and minimax properties of Chebyshev polynomials within each transformation layer, the Cheby-KAN effectively extracts complex patterns from input data, significantly enhancing the model's expressive capacity and predictive accuracy.
\subsection{Hybrid deep learning architecture}
To achieve efficient and accurate prediction of the two-dimensional airfoil pressure field, this study proposes a hybrid deep learning architecture that integrates geometric feature extraction with flow field mapping.
\subsubsection{Geometric feature extraction module}
\label{sec:sec2.2.1}
\begin{figure} 
    \centering
    \includegraphics[width=8cm]{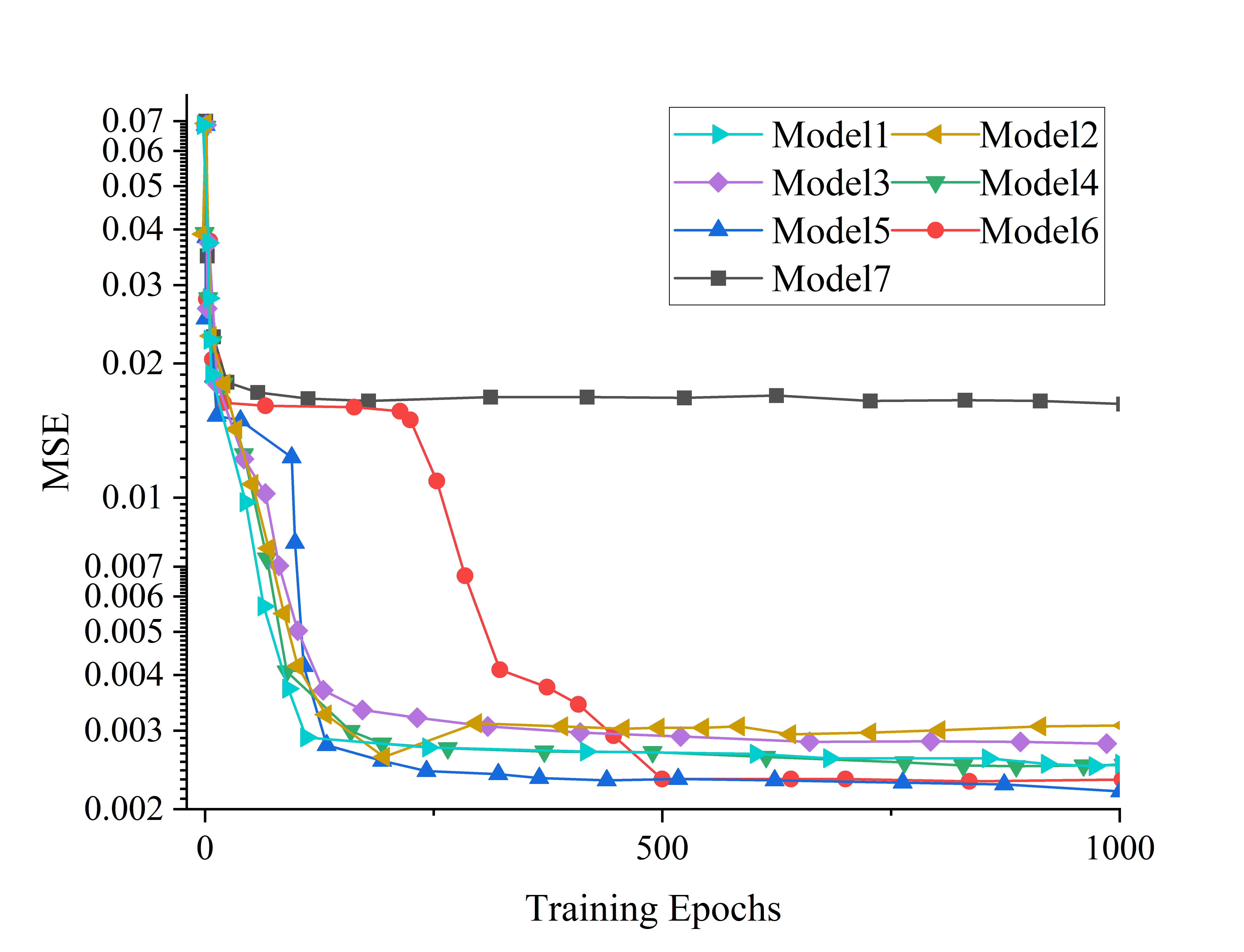}
    \caption{Comparison of MSE after network training with different structures}
    \label{fig:Fig5}
\end{figure}

\begin{figure} 
    \centering
    \includegraphics[width=14cm]{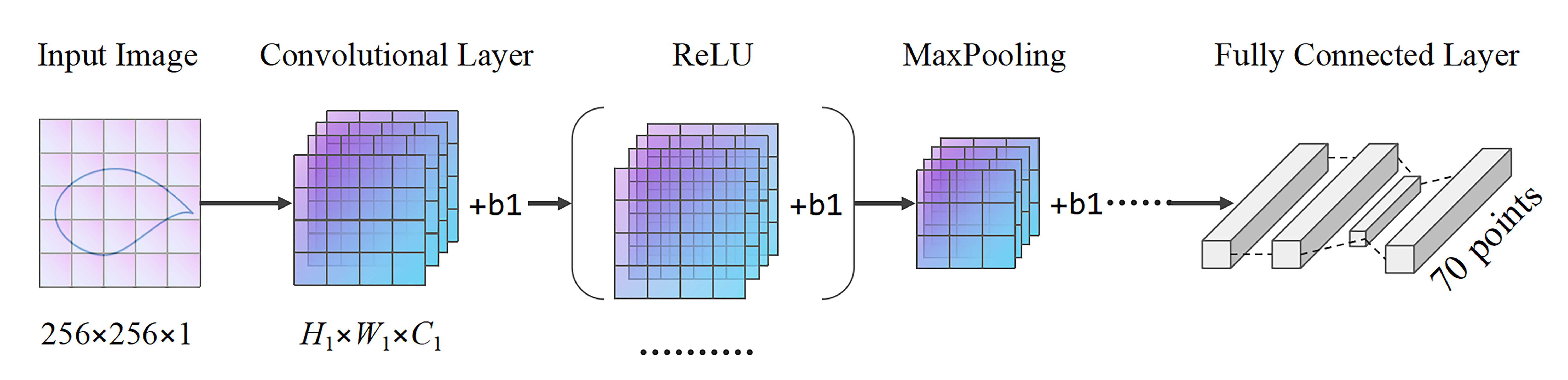}
    \caption{Schematic diagram of grid structure for geometric parameterization of airfoil}
    \label{fig:Fig6}
\end{figure}

The geometric feature extraction module employs a deep CNN to perform parametric modeling of airfoil geometries (Figure \ref{fig:Fig6}). In this part, 1,549 heterogeneous airfoils were selected from the UIUC database, including airfoils from the NACA series, supercritical airfoils, and natural laminar flow airfoils. Each airfoil contour is fitted with a cubic B-spline curve, uniformly sampled, and normalized into a standardized grayscale image. These images are then fed into a lightweight CNN encoder. The encoder comprises four blocks, each consisting of a convolutional layer (with a $4 \times 4$ kernel and a stride of 1) followed by a max-pooling layer (with a $3 \times 3$ window). Table \ref{tab:Tab1} presents the specific configuration of the convolution module section. Through this process, the spatial dimensions are progressively reduced, and the features are ultimately mapped to a 16-dimensional latent vector via a fully connected layer. This 16 dimensional latent vector can be restored to 70 data points of the original airfoil through the fully connected layer. The model was trained for 1,500 epochs using an initial learning rate of 0.000375. By comparing different fully connected configurations (Table \ref{tab:Tab2}), experimental results indicate that the Model 5 architecture achieves the best performance on the test set (Figure \ref{fig:Fig5}).

\begin{table}
 \caption{Convolutional neural network architecture}
  \centering
  \begin{tabular}{llllllll}
    \toprule
    \cmidrule(r){1-2}
    Layer &	Type &	Input Channel &	Output Channel &	Convolution Kernel Size &	Step &	Filling &	Output Shape\\
    \midrule
    1 &	Input Layer &	- &	- &	- &	- &	-	& $256 \times 256 \times 1$\\
    2 & Conv2d & 1 & 32 & $4\times4$ & 1 & 1 & $255\times255\times32$ \\
    3 & MaxPool2d & - & - & $3\times3$ & 3 & 0 & $85\times85\times32$ \\
    4 & Conv2d & 32 & 32 & $4\times4$ & 1 & 1 & $84\times84\times32$ \\
    5 & MaxPool2d & - & - & $3\times3$ & 3 & 0 & $28\times28\times32$ \\
    6 & Conv2d & 32 & 64 & $4\times4$ & 1 & 1 & $27\times27\times64$ \\
    7 & MaxPool2d & - & - & $3\times3$ & 3 & 0 & $9\times9\times64$ \\
    8 & Conv2d & 64 & 64 & $4\times4$ & 1 & 1 & $8\times8\times64$ \\
    9 & MaxPool2d & - & - & $3\times3$ & 3 & 0 & $2\times2\times64$ \\
    \bottomrule
  \end{tabular}
  \label{tab:Tab1}
\end{table}

\begin{table}
 \caption{Different fully connected layer structures and MSEs}
  \centering
  \begin{tabular}{lllllll}
    \toprule
    \cmidrule(r){1-2}
    Name &	Hidden Layer1 &	Hidden Layer2 &	Hidden Layer3 &	Output Layer &	Train MSE &	Test MSE\\
    \midrule
    Model1 & $6\times256$ & $1\times16$ & $6\times256$ & 70 & 0.002136 & 0.004814 \\
    Model2 & $7\times256$ & $1\times16$ & $7\times256$ & 70 & 0.003128 & 0.007172 \\
    Model3 & $8\times256$ & $1\times16$ & $8\times256$ & 70 & 0.002761 & 0.009777 \\
    Model4 & $9\times256$ & $1\times16$ & $9\times256$ & 70 & 0.002943 & 0.004046 \\
    Model5 & $10\times256$ & $1\times16$ & $10\times256$ & 70 & 0.002218 & 0.001966 \\
    Model6 & $11\times256$ & $1\times16$ & $11\times256$ & 70 & 0.002247 & 0.005418 \\
    Model7 & $12\times256$ & $1\times16$ & $12\times256$ & 70 & 0.016310 & 0.037177 \\
    \bottomrule
  \end{tabular}
  \label{tab:Tab2}
\end{table}

\subsubsection{Flow field prediction module}
The flow field prediction module employs a Chebyshev polynomial-based neural network to construct an end-to-end mapping from inputs to the output pressure field. The low-dimensional geometric feature vectors of airfoils are extracted using a CNN-based encoder and subsequently normalized. This global feature vector, together with spatial coordinates within the flow domain and flight condition parameters—such as Reynolds number, angle of attack, and Mach number—forms the input to the network. The model outputs the normalized pressure values at the corresponding spatial locations under the specified flow conditions.

To implement the Cheby-KAN architecture, a custom Cheby-KAN layer is defined that integrates a linear transformation with a Chebyshev polynomial transformation. Key implementation steps of the Cheby-KAN layer are as follows:
\begin{enumerate}
  \item Initialization: The base weight matrix $W_{j,i}^{BasicL}$, Chebyshev coefficients $c_{j,i,k}$, and bias vector $b$ are initialized. A nonlinear activation function $\sigma(x)$ (e.g., SiLU) is also defined.
  \item Forward Propagation: During forward pass, the input first undergoes a base linear transformation. The result is then passed through a Chebyshev polynomial transformation and summed with the original linear output. A bias term is added to produce the final output.Specifically, the forward propagation process can be described as:
    \begin{equation}
      {output} = {BaseLinear}(\sigma(x)) + \sum_{k=0}^{7} c_n \cdot T_n(x) + b
    \end{equation}
    Where $\sigma(x)$ denotes the SiLU activation function, $T_n(x)$ denotes the n-th order Chebyshev polynomial, $c_n$ denotes the Chebyshev coefficient, and $b$ denotes the bias term. The transformation employs Chebyshev polynomials up to the 7-th order.
  \item Initialization and Regularization:In terms of parameter initialization, the basic weights are initialized using the Kaiming uniform initialization strategy, while the Chebyshev coefficients are initialized using a uniform distribution with a standard deviation of:
    \begin{equation}
      \sigma = \frac{c_s}{\sqrt{d_n}}
    \end{equation}
    Where $\sigma$ denotes the standard deviation of Chebyshev coefficient initialization, $c_s$ denotes the Chebyshev coefficient scaling factor (default value is 1.0), $d_n$ denotes the dimension of input features. To control model complexity and prevent overfitting, we introduced the $L_2$ regularization term of Chebyshev coefficient in the loss function. The total loss function is defined as the sum of mean square error and regularization term, with the regularization coefficient set to 1.0 by default. During the training process, a global gradient pruning strategy is adopted to limit the gradient norm to within 1.0, ensuring the stability of the training process.
\end{enumerate}

The total loss function consists of a data fitting term and a regularization term:
\begin{equation}
  L_{total} = \frac{1}{N} \sum_{i=1}^{N}(y_i - \hat{y}_i)^2 + \lambda \cdot \frac{1}{L} \sum_{l=1}^{L} \frac{a}{d_{out} d_{in} (K+1)} \sum_{o=1}^{d_{out}} \sum_{i=1}^{d_{in}} \sum_{k=0}^{K} (c_{l,o,i,k})^2
\end{equation}
Where $N$ denotes batch size, $y_i$ and $\hat{y}_i$ denote real value and predicted value, $L$ denotes network layer number, $d_{out}$ and $d_{in}$ denote output and input feature dimensions, $K$ denotes Chebyshev polynomial order, $c_{l,o,i,k}$ denotes Chebyshev coefficient of the l-th layer, $\lambda$ denotes Regularization coefficient (default 1.0).

The Cheby-KAN neural network consists of an input layer, multiple hidden layers, and an output layer. The input layer receives a total of 20 parameters, including spatial coordinates $(x,y)$, flow condition parameters such as Reynolds number $(Re)$, angle of attack $(\alpha)$, as well as the 16-dimensional airfoil feature vector extracted by the CNN encoder. The network contains 6 hidden layers with a structural layout of $[128,256,512,256,128,64]$, meaning the number of grid points per layer follows this sequence. The output layer consists of a single neuron that produces the final scalar output: the pressure value.

\begin{figure} 
    \centering
    \includegraphics[width=8cm]{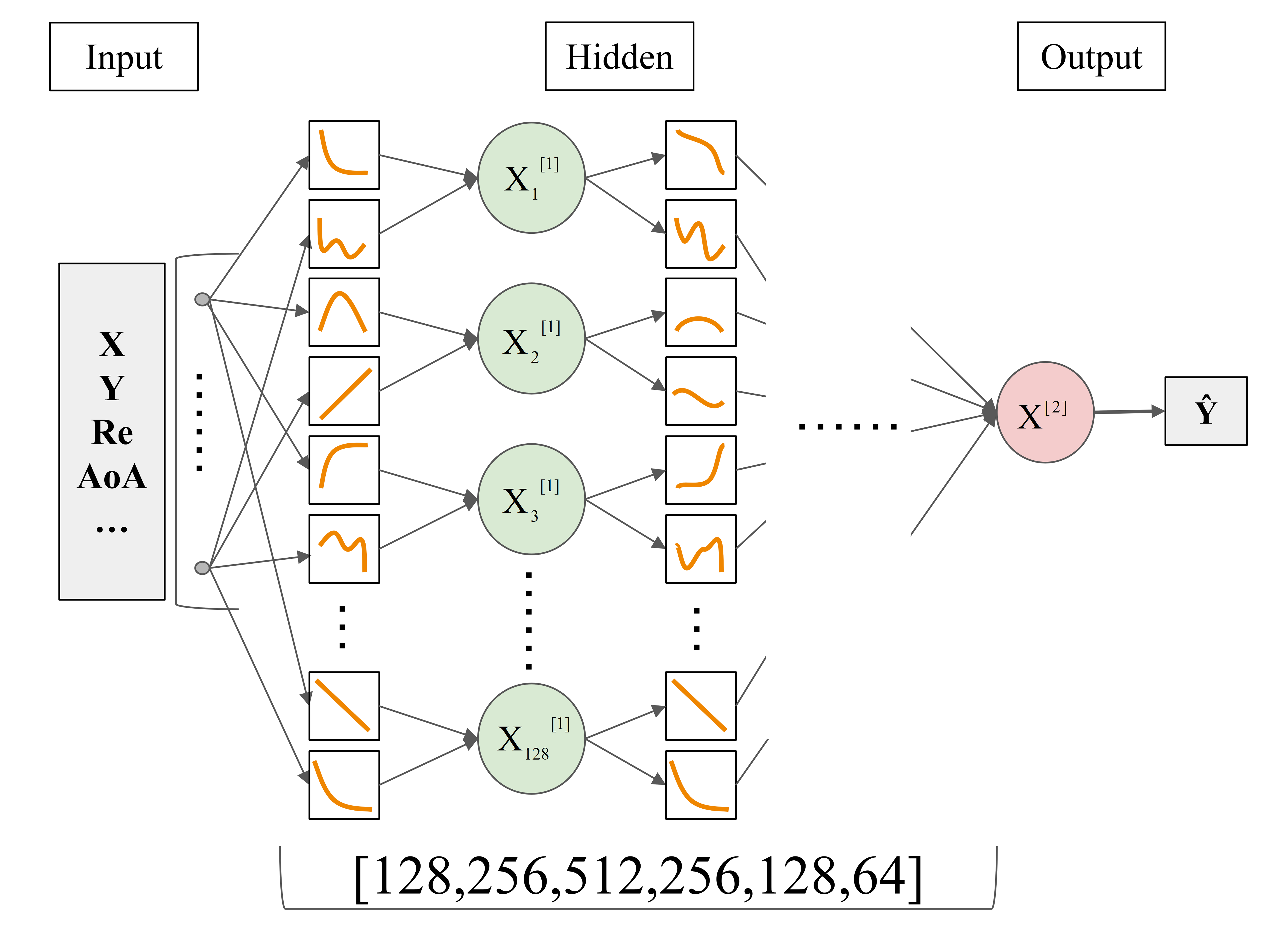}
    \caption{Cheby-KAN neural network structure diagram}
    \label{fig:Fig7}
\end{figure}

The complete architecture of the Cheby-KAN model is illustrated in Figure \ref{fig:Fig7}.

As a benchmark comparison, we briefly introduce the flow field prediction architecture based on Multilayer Perceptron (MLP). The fundamental computational unit of MLP is the neuron (Figure \ref{fig:Fig3}), and its mathematical model can be abstracted as:
\begin{equation}
  h = f(\sum_{i=1}^{n} w_i x_i + b)
\end{equation}
MLP achieves hierarchical feature extraction by introducing hidden layers. Each hidden layer contains multiple neurons and receives outputs from all neurons in the preceding layer through fully-connected connections. The activation value of the j-th neuron in the l-th hidden layer is computed as:
\begin{equation}
  h_j^{(l)} = f(\sum_{i=1}^{n} w_{j,i}^{(l)} h_j^{(l-1)} + b_j^{(l)})
\end{equation}
Where $h_j^{(l-1)}$ denotes the output from the previous layer, $w_{j,i}^{(l)}$ is the weight matrix, and $b_j^{(l)}$ is the bias term. Through successive nonlinear transformations, the hidden layers progressively extract higher-level flow field features, such as boundary layer separation and vortex interference.

While maintaining identical input configurations, we designed three MLP architectures and implemented a dynamic learning rate scheduling strategy with an initial learning rate of $5 \times 10^{-5}$, decaying by a factor of $\lambda = 0.8$ every 20 training steps. Table \ref{tab:Tab3} summarizes the structural characteristics and training errors of the different MLP configurations. Table \ref{tab:Tab4} compares the prediction errors and computational times of the CFD solver, MLP-based prediction module, and Cheby-KAN prediction module on an RTX 4070 GPU. It is noteworthy that the standard Kolmogorov–Arnold Network (KAN), without Chebyshev polynomial enhancement, struggles to handle the substantial computational load presented by the dataset. In contrast, the Cheby-KAN achieves significantly improved prediction accuracy at the cost of increased training time.

\begin{table}
 \caption{The structural characteristics and training errors of different MLP models}
  \centering
  \begin{tabular}{llllll}
    \toprule
    \cmidrule(r){1-2}
    Name &	Input Layer &	Hidden Layer &	Output Layer &	Test MSE &	Train MSE\\
    \midrule
    MLP-1 & $1\times25$ & $7\times256$ & $1\times1$ & 0.00001490 & 0.00010106 \\
    MLP-2 & $1\times25$ & $9\times256$ & $1\times1$ & 0.00004848 & 0.00009586 \\
    MLP-3 & $1\times25$ & $11\times256$ & $1\times1$ & 0.00000997 & 0.00013954 \\
    \bottomrule
  \end{tabular}
  \label{tab:Tab3}
\end{table}

\begin{table}
 \caption{Comparison of model time under the same computational hardware}
  \centering
  \begin{tabular}{lllll}
    \toprule
    \cmidrule(r){1-2}
    Model &	\makecell{Time for \\Training} &	\makecell{Time for Prediction \\ (Single Condition)} &	\makecell{Time for Prediction \\ (100 Conditions)} &	\makecell{Mean Relative Error \\ (100 Conditions)}\\
    \midrule
    CFD & -- & $>30$ min & $>3000$ min & 0\% \\
    MLP & $\sim 1340$ min & $\sim 3.4$ s & $\sim 350$ s & $\sim 2.5\%$ \\
    Cheby-KAN & $\sim 7860$ min & $\sim 7.9$ s & $\sim 800$ s & $\sim 1.3\%$ \\
    \bottomrule
  \end{tabular}
  \label{tab:Tab4}
\end{table}

\section{Data preparation}
In the UIUC database, airfoil geometry data are provided as discrete coordinate points. To render these data suitable for CNN input, a standardized processing pipeline was established.

During geometric parameterization, smoothing the airfoil contour and applying appropriate interpolation are critical steps. In this study, a cubic uniform B-spline curve was used to fit the original discrete points, effectively filtering out high-frequency noise and ensuring contour smoothness. Furthermore, 70 points were equidistantly sampled from the leading edge to the trailing edge to obtain a uniform and continuous geometric representation for subsequent analysis.

To eliminate scale variations among different airfoils and mitigate adverse effects on model training, the coordinate data were normalized using the following equations:
\begin{equation}
  x_{norm} = \frac{x - x_{min}}{x_{max} - x_{min}}
\end{equation}
\begin{equation}
  y_{norm} = \frac{y - y_{min}}{y_{max} - y_{min}}
\end{equation}
Where $x_{min}$ and $x_{max}$ represent the minimum and maximum values of the airfoil chord coordinates, respectively, and $y_{avg}$ denotes the average thickness.

\begin{figure} 
    \centering
    \includegraphics[width=8cm]{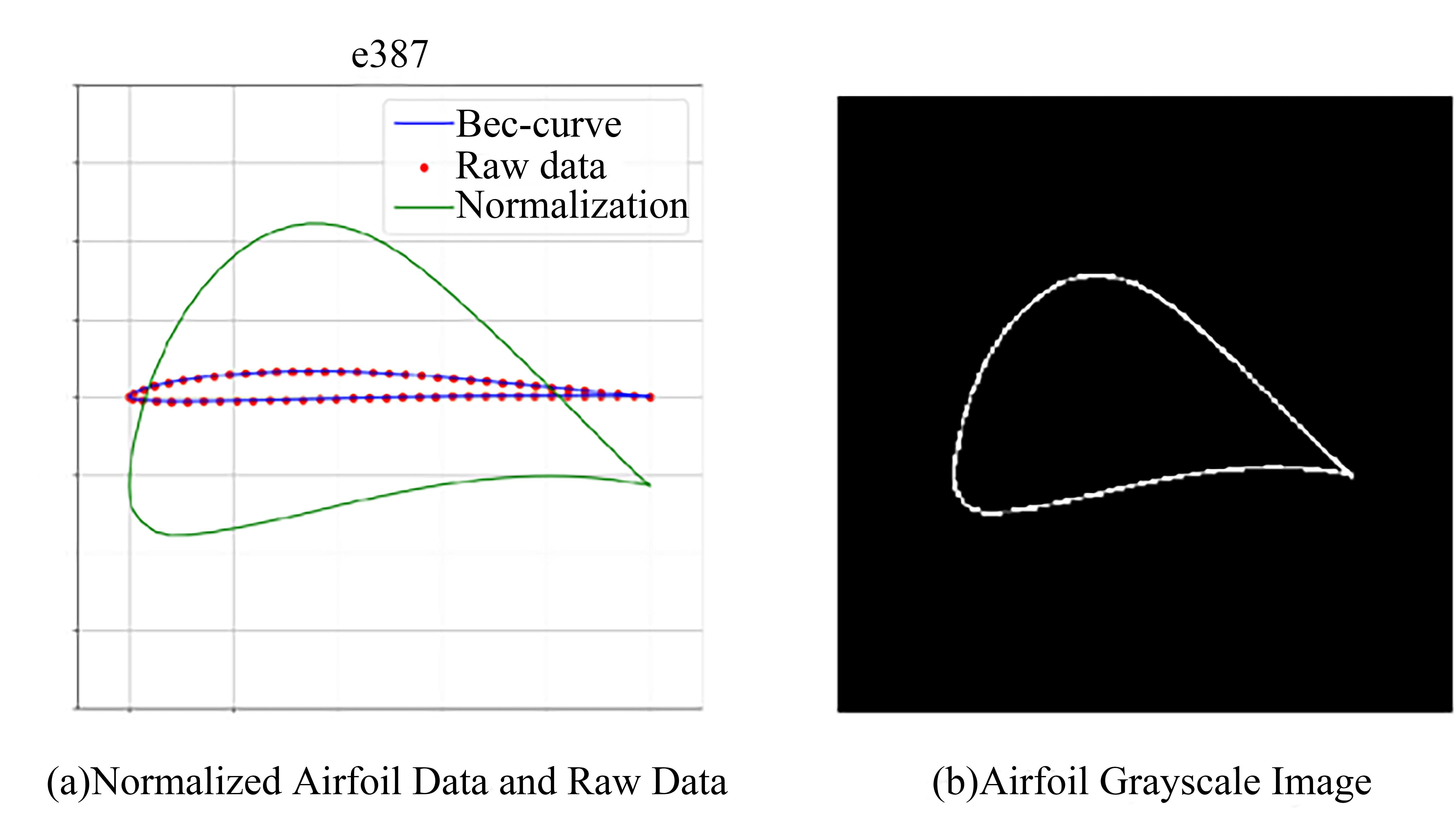}
    \caption{Data processing results of airfoil (e387)}
    \label{fig:Fig8}
\end{figure}

After normalization, all airfoil profiles are mapped to the interval $[-1,1]$, effectively eliminating dimensional influences while preserving their intrinsic geometric characteristics Figure (\ref{fig:Fig8}). Subsequently, using the Matplotlib library, each normalized airfoil contour was converted into a grayscale image in which pixel values corresponding to the airfoil profile were assigned a value of 1 (white), and the background pixels were set to 0 (black) (Figure \ref{fig:Fig8}).

To acquire the flow field data required for training, this study selected several widely-recognized airfoils—including the RAE2822, NACA0012, e387, and mh38—as the dataset. These airfoils are well-established benchmarks in aerodynamic studies, with their flow characteristics extensively documented in the literature\cite{selig2004wind}\cite{shin1992results}\cite{mcghee1988experimental}, making them suitable for validating our CFD setup and the proposed model. After analyzing computational accuracy, cost, and grid independence, a mesh of approximately 2 million elements per airfoil was determined to ensure sufficient accuracy. Consequently, the ANSYS Fluent solver was selected as the computational fluid dynamics tool to guarantee that every case exceeded this mesh count. The SST $K-\omega $ turbulence model was employed for closure. Boundary conditions included a far-field pressure inlet with a static pressure of 101,325 Pa and a temperature of 288.15 K. All airfoils had a characteristic chord length of 0.15 m, and the incoming flow velocity varied from 40 m/s to 80 m/s.

For each airfoil, 83 distinct operating conditions were randomly sampled from predefined ranges (Reynolds number: $4.05 \times 10^5 - 8.23 \times 10^5$, angle of attack: $0^\circ - 8^\circ $). The resulting dataset was partitioned into 90\% for training and 10

To structure the CFD results for neural network input, a coordinate transformation was applied to the flow field data surrounding each airfoil. Specifically, a square domain measuring $[-1.8,1.8] \times [-1.8,1.8]$ was defined around the airfoil to capture the near-field region. A structured grid was generated within this domain to sample flow variables. The extracted data were normalized using the following formula, ultimately forming the pressure distribution cloud map (Figure \ref{fig:Fig8}):
\begin{equation}
  p_{norm} = \frac{p - p_{min}}{p_{max} - p_{min}}
\end{equation}
Where $p_{norm}$ denotes the normalized pressure, and $p_{min}$ and $p_{max}$ represent the minimum and maximum values of the sectional pressure data, respectively.
\section{Results and analysis}
This study employed the CUDA platform and the PyTorch framework for model training. The key training hyperparameters are summarized in Table \ref{tab:Tab5}.

\begin{table}
 \caption{Training hyperparameters for the Cheby-KAN network}
  \centering
  \begin{tabular}{llllll}
    \toprule
    \cmidrule(r){1-2}
    Batchsize	& Epoch	& \makecell{Chebyshev Polynomial \\ Order} &	\makecell{Maximum Norm of \\ Gradien Clipping}  &	\makecell{Initial Learning \\ Rate} &	Gamma\\
    \midrule
    32 &	500	& 7 &	1.0 &	0.00005 &	0.4\\
    \bottomrule
  \end{tabular}
  \label{tab:Tab5}
\end{table}

The hybrid CNN-Cheby-KAN architecture was trained using the Adam optimizer with a dynamic learning rate scheduler (StepLR with $gamma = 0.4$). The model was saved after each epoch to maximize interpretability of the KAN component and facilitate detailed analysis of training progression. Mean squared error (MSE) was used as the loss function, while mean absolute error (MAE) and the coefficient of determination ($R^2$) were adopted as evaluation metrics to assess model performance.
\begin{equation}
  MSE = \frac{1}{n} \sum_{i=1}^{n} (p_{pre} - p_{tru})^2
\end{equation}
\begin{equation}
  MAE = \frac{1}{n} \sum_{i=1}^{n} \left\lvert p_{pre}^{(i)} - p_{tru}^{(i)} \right\rvert
\end{equation}
\begin{equation}
  R^2 = 1 - \frac{\sum_{i=1}^{n} (p_{pre} - p_{tru})^2}{\sum_{i=1}^{n} (p_{tru} - \bar{p}_{tru})^2}
\end{equation}
where $p_{pre}$ denotes the predicted pressure value, $p_{tru}$ represents the CFD-computed value, and $\bar{p}_{tru}$ the mean of the CFD-generated pressure data. This section analyzes the predictive accuracy, spanwise interpolation capability, and error distribution characteristics of the model under representative operating conditions.

Owing to the structural characteristics of the KAN network—where learnable univariate functions replace traditional linear weights—the computational and memory requirements are significantly higher compared to conventional architectures. This increased resource consumption motivated us to systematically investigate the influence of dataset size on model performance, aiming to identify a cost-effective training strategy that balances accuracy and computational expense.

\begin{figure} 
    \centering
    \includegraphics[width=8cm]{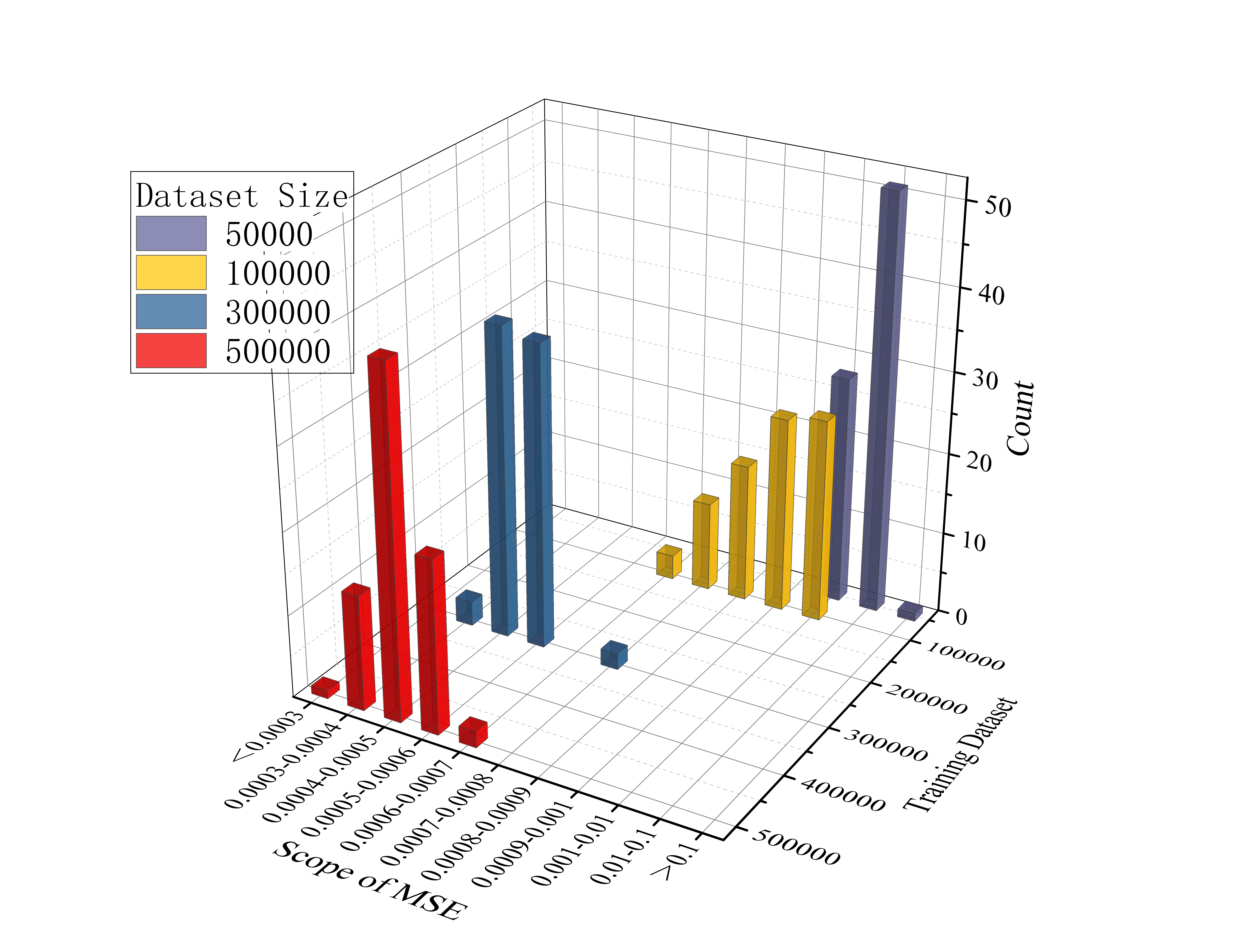}
    \caption{Test MSE comparison of training models using different batches of data sets}
    \label{fig:Fig9}
\end{figure}

To investigate the impact of dataset size on model performance, we used four different dataset sizes, including 50,000, 100,000, 300,000, and 500,000 samples. These datasets were randomly extracted from the complete flow field dataset. We trained models on these datasets using identical training parameters in the same training environment, resulting in four distinct models. The predictive accuracy of each model was evaluated on 80 flow field conditions, and the corresponding MAE values were computed. The distribution of MAE values for each model is shown in Figure \ref{fig:Fig9} and Table \ref{tab:Tab4}.

\begin{table}
 \caption{Comparison of MSE testing between training models using datasets from different batches}
  \centering
  \begin{tabular}{lllll}
    \toprule
    \cmidrule(r){1-2}
    Dataset Size & 50,000 & 100,000 & 300,000 & 500,000 \\
    \midrule
    $<0.0003$ & 0 & 0 & 0 & 1 \\
    $0.0003-0.0004$ & 0 & 0 & 3 & 14 \\
    $0.0004-0.0005$ & 0 & 0 & 38 & 42 \\
    $0.0005-0.0006$ & 0 & 0 & 37 & 21 \\
    $0.0006-0.0007$ & 0 & 3 & 0 & 2 \\
    $0.0007-0.0008$ & 0 & 11 & 2 & 0 \\
    $0.0008-0.0009$ & 0 & 17 & 0 & 0 \\
    $0.0009-0.001$ & 0 & 24 & 0 & 0 \\
    $0.001-0.01$ & 28 & 25 & 0 & 0 \\
    $0.01-0.1$ & 51 & 0 & 0 & 0 \\
    $>0.1$ & 1 & 0 & 0 & 0 \\
    \bottomrule
  \end{tabular}
  \label{tab:Tab6}
\end{table}

The results demonstrate that dataset size significantly influences prediction accuracy. In terms of mean MAE, the model trained on 50,000 samples exhibited the highest error, with most values distributed above 0.001. Performance improved with 100,000 samples, yielding a lower average MAE. A further reduction in error was observed with 300,000 samples, where the average MAE decreased and most values fell between 0.0004 and 0.0006.

However, when the dataset was increased to 500,000 samples, the average MAE did not decrease significantly relative to the 300,000-sample model, indicating a potential performance plateau. This suggests that the marginal benefit of adding more data diminishes beyond a certain scale, and the relationship between dataset size and accuracy becomes nonlinear. Consequently, merely increasing data volume cannot further enhance model performance once this saturation point is reached.

The experimental results indicate that the model achieves strong generalization capability and high predictive accuracy when trained on a dataset of 300,000 samples. Furthermore, increasing the dataset size substantially increases per-epoch training time. Therefore, this study employs the 300,000-sample dataset for all subsequent in-depth analysis.In the case of a large number of data samples, we believe that such a data set pruning is necessary for Cheby-Kan networks.

\begin{figure} 
    \centering
    \includegraphics[width=8cm]{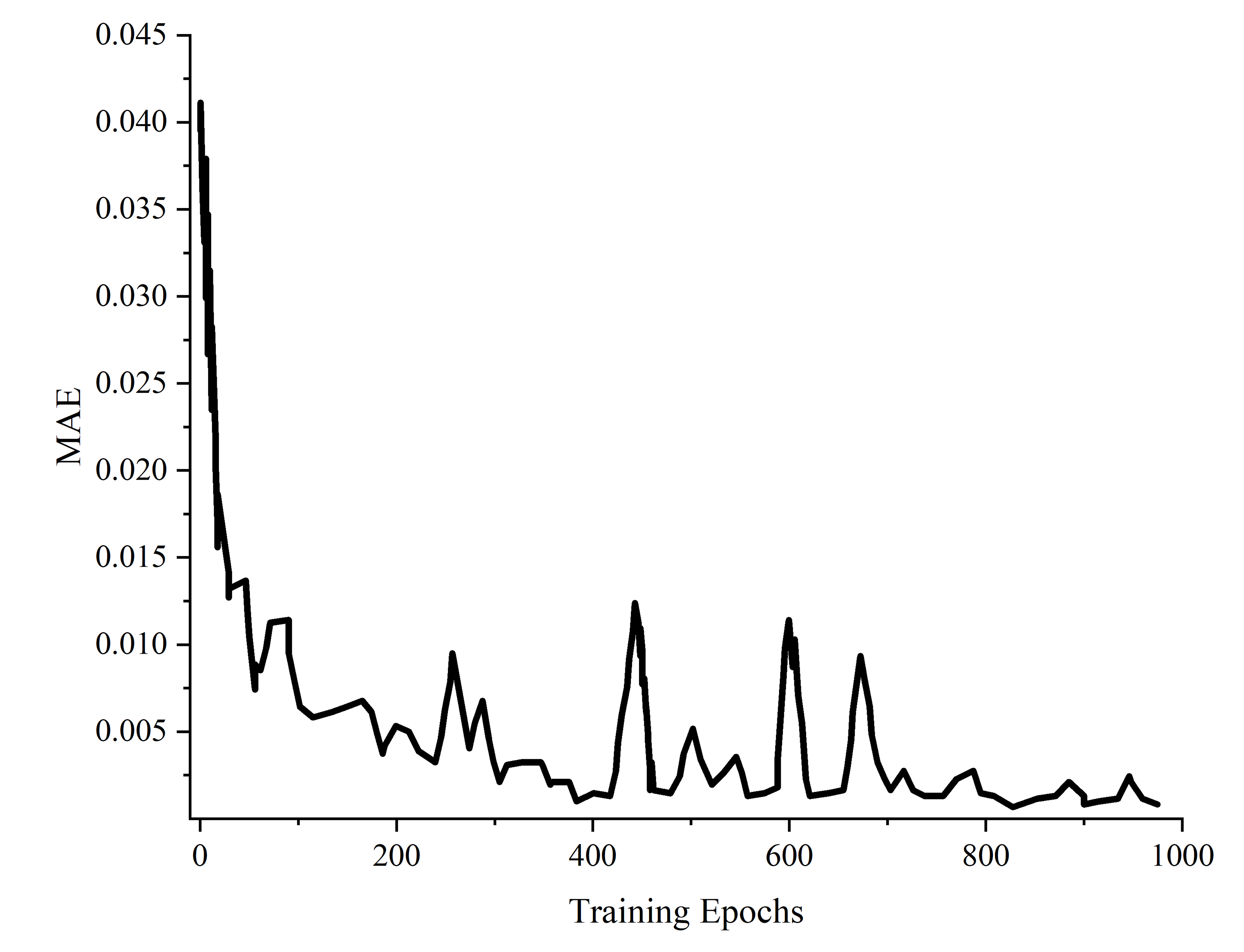}
    \caption{MAE curve of 1000 rounds under 300000 training sets}
    \label{fig:Fig10}
\end{figure}

For this dataset, the training duration was extended to 1,000 epochs to further enhance model accuracy. The training set size, batch size, gamma value, and other hyperparameters remained unchanged. The model was evaluated on a held-out test set, and the progression of MAE over training epochs is illustrated in Figure \ref{fig:Fig10}.

The results show that the MAE decreased rapidly during the initial phase of training and stabilized around 0.005 after approximately 200 epochs, though occasional oscillations were observed. The amplitude and frequency of these oscillations decreased as training progressed. This behavior can be attributed to the relatively high initial learning rate under the decay strategy, which initially introduced instability and fluctuations in errors. As the learning rate gradually decreased, the training process became more stable. However, the overall reduction in MAE was limited, especially beyond around 780 epochs, at which point the MAE converged to approximately 0.001. Overall, the model achieved satisfactory accuracy, and the version trained for 1,000 epochs was selected for final performance evaluation.

\begin{figure} 
    \centering
    \includegraphics[width=14cm]{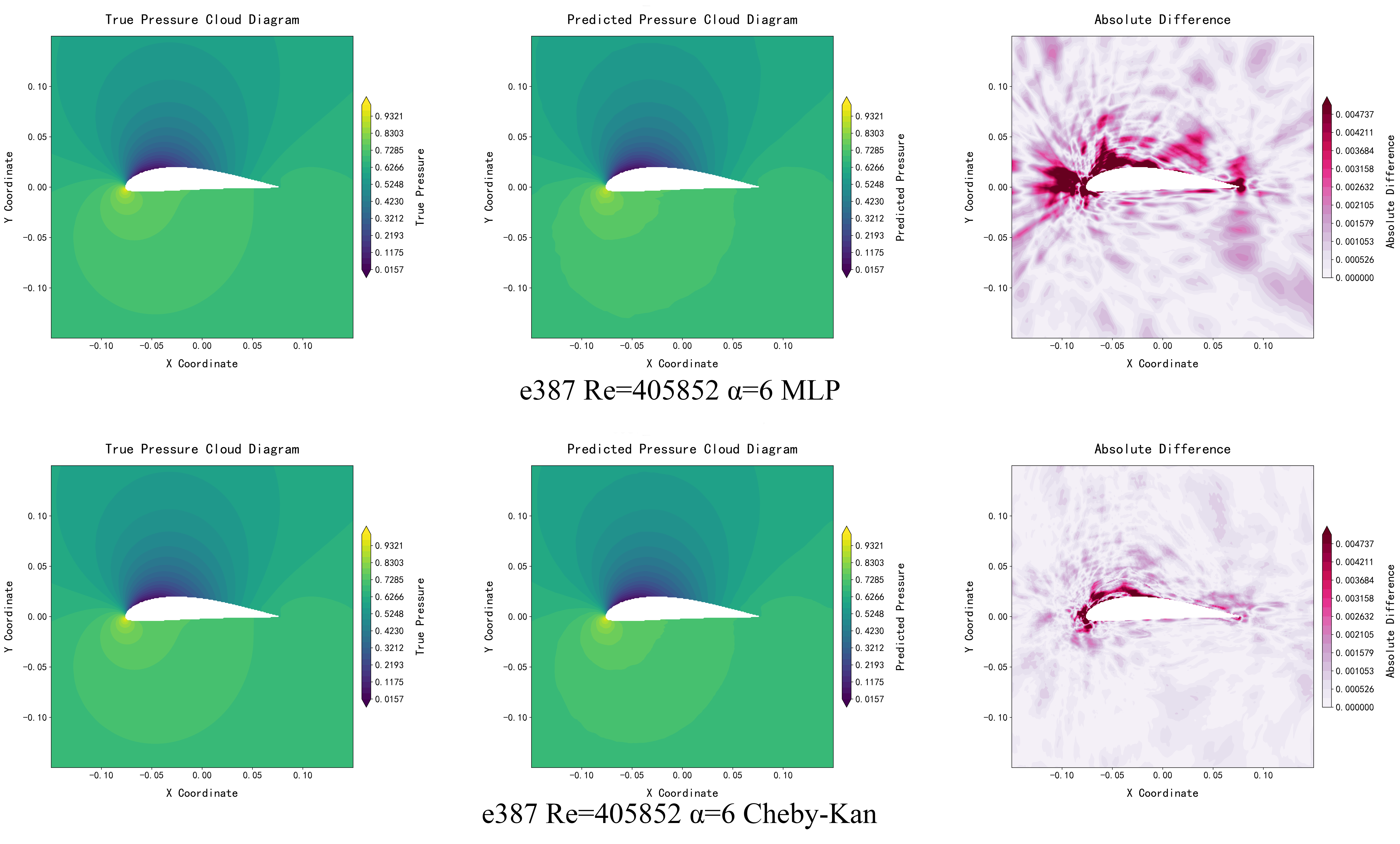}
    \caption{Comparison of prediction results between two models under the same operating conditions}
    \label{fig:Fig11}
\end{figure}

To validate the superior performance of the proposed Cheby-KAN model, a conventional Multilayer Perceptron (MLP) was also trained under identical conditions for comparative analysis. Its predictive results were compared with those of Cheby-KAN, as illustrated in Figure \ref{fig:Fig11}.

\begin{figure} 
    \centering
    \includegraphics[width=14cm]{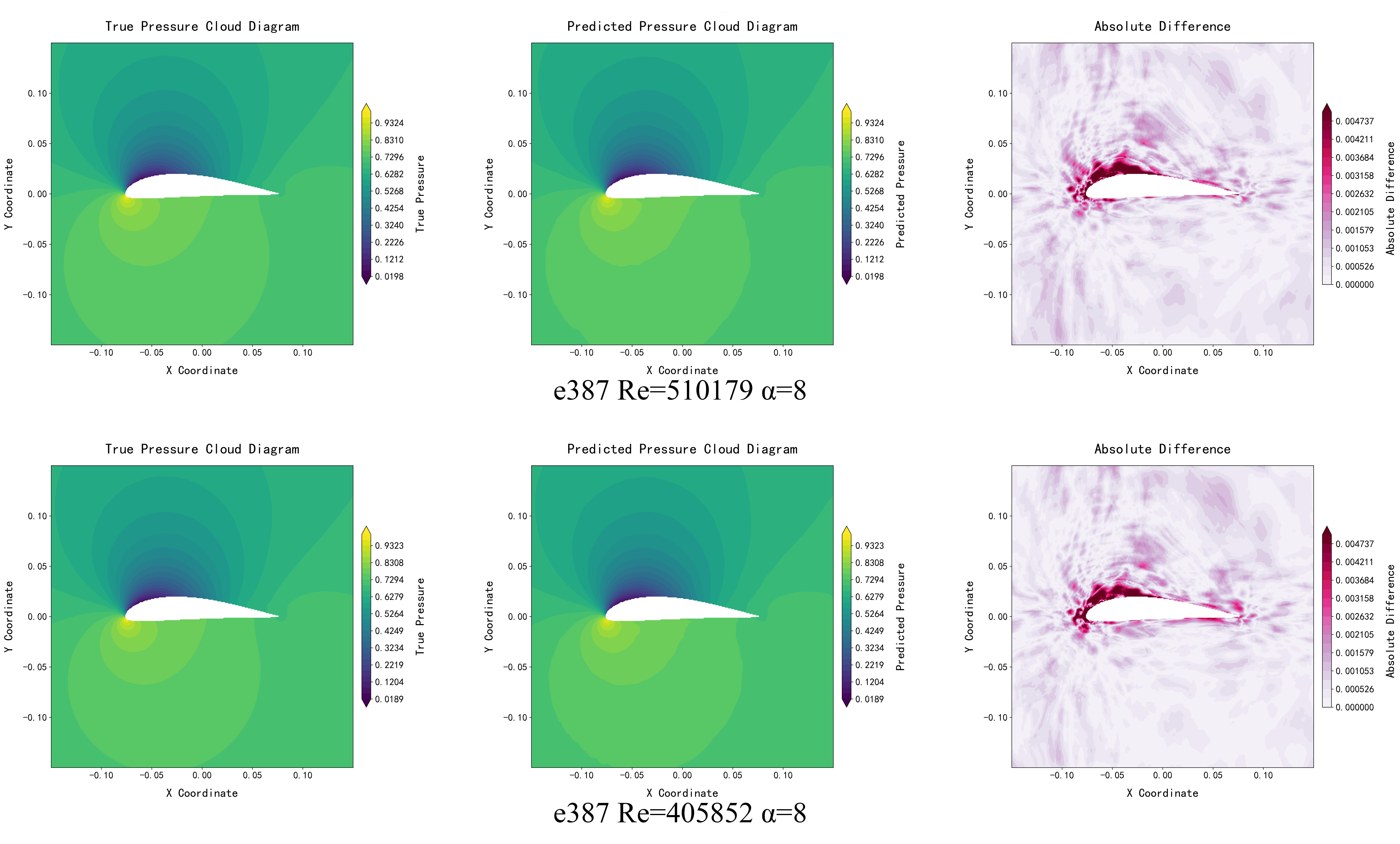}
    \caption{Prediction results of e387 airfoil using Cheby KAN model}
    \label{fig:Fig12}
\end{figure}

\begin{figure} 
    \centering
    \includegraphics[width=14cm]{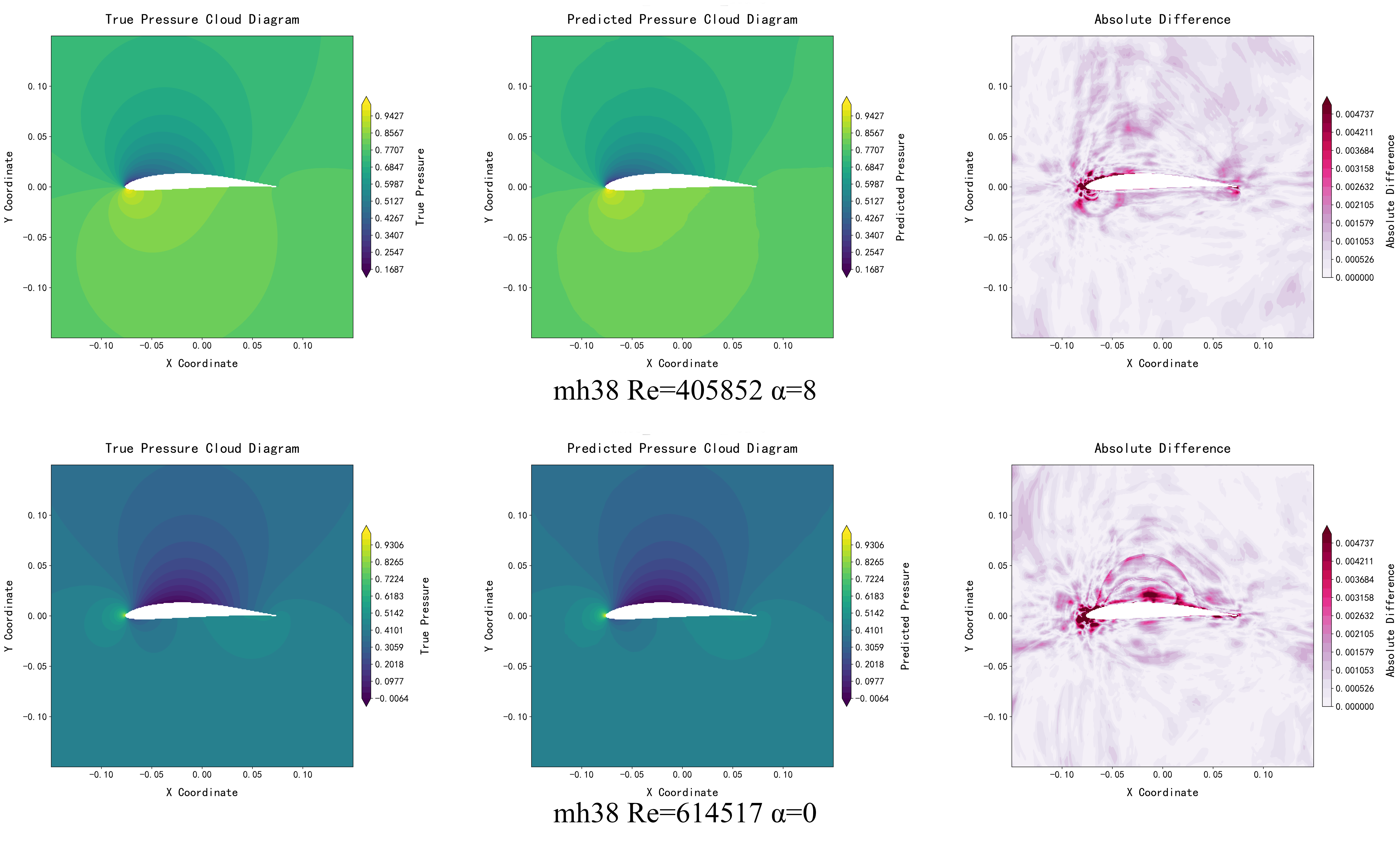}
    \caption{Prediction results of mh38 airfoil using Cheby KAN model}
    \label{fig:Fig13}
\end{figure}

\begin{figure} 
    \centering
    \includegraphics[width=14cm]{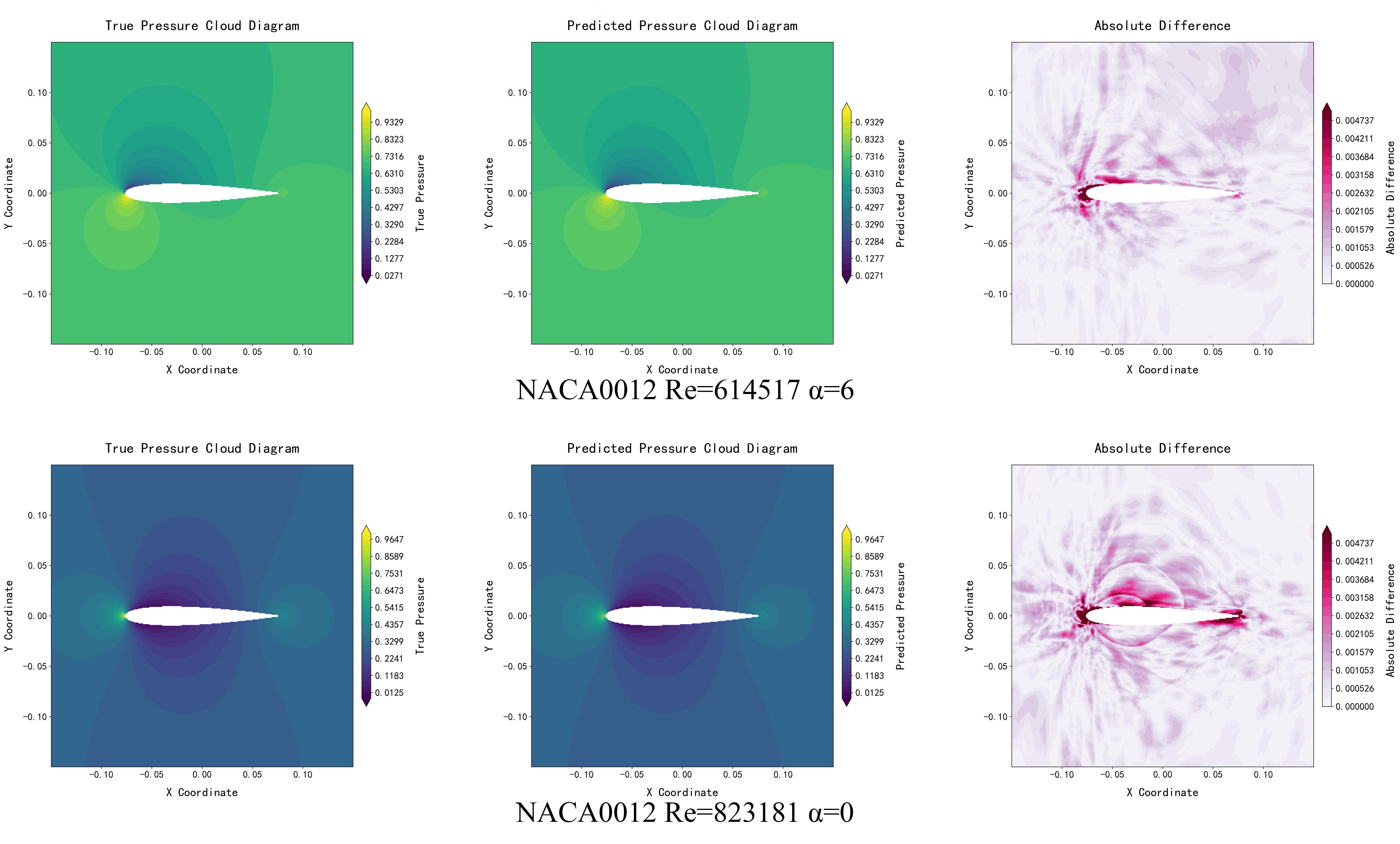}
    \caption{Prediction results of NACA0012 airfoil using Cheby KAN model}
    \label{fig:Fig14}
\end{figure}

The results indicate that, due to its structural limitations, the MLP struggled to effectively capture complex underlying data relationships—even when supplied with a sufficient amount of training data. In contrast, Cheby-KAN demonstrated a strong capacity for modeling both intricate linear and nonlinear dependencies, leading to significantly improved predictive accuracy. To achieve a level of accuracy comparable to Cheby-KAN in predicting two-dimensional airfoil pressure distributions, the MLP would require a substantially larger dataset. Additional prediction results of the Cheby-KAN model under various operating conditions are presented in Figure \ref{fig:Fig12} through Figure \ref{fig:Fig14}.

These test conditions cover the performance of various common airfoils at different Reynolds numbers and angles of attack ($\alpha$), covering the main subsonic flow states. The results confirm that the model achieves high predictive accuracy across these scenarios. Performance is notably strong in the far-field region surrounding the airfoil; however, prediction errors remain relatively elevated within the boundary layer for certain flow configurations.

\begin{figure} 
    \centering
    \includegraphics[width=14cm]{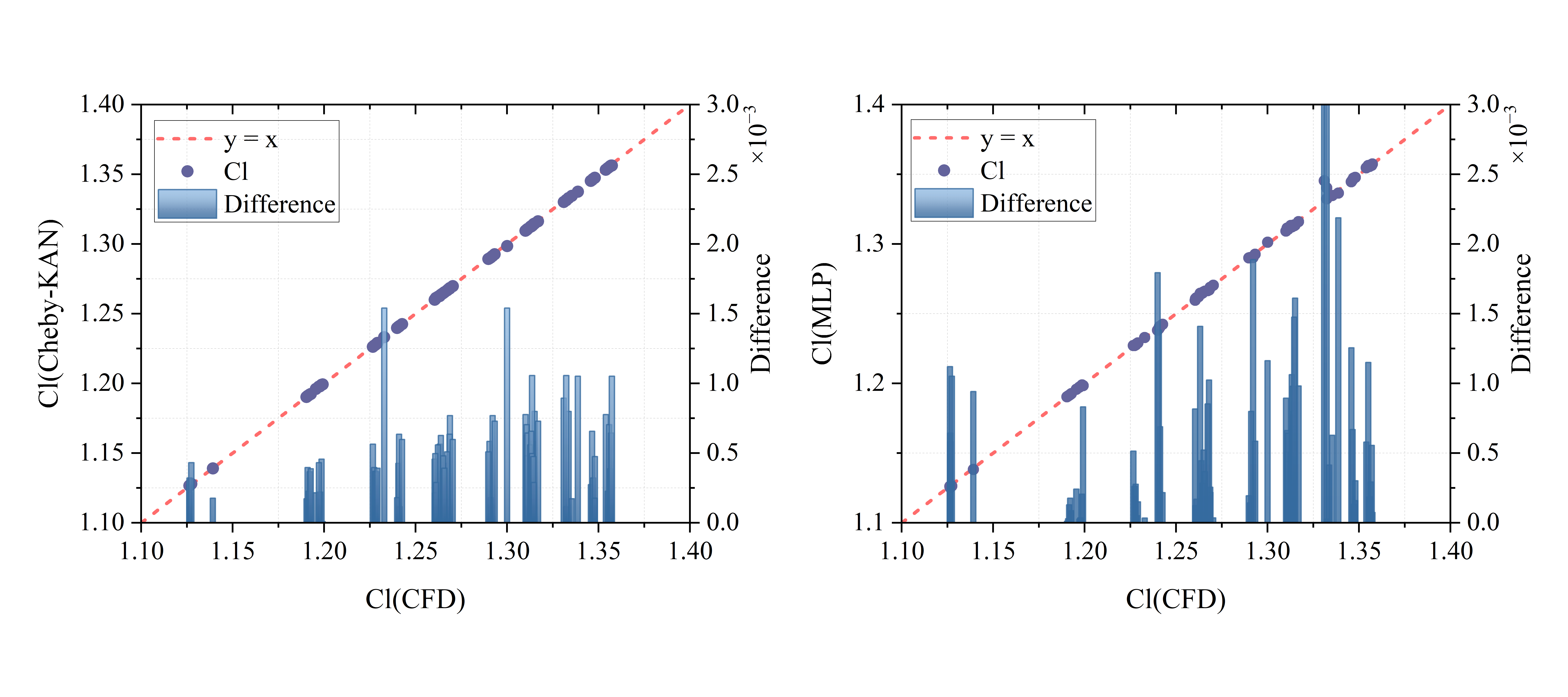}
    \caption{Comparison of lift coefficients and errors between different models}
    \label{fig:Fig15}
\end{figure}

Figure \ref{fig:Fig15} depicts the predicted lift coefficients for all airfoils, derived from the predicted pressure distributions, with the bar graph representing the discrepancy between the predicted lift coefficients and the CFD results. Although both methods achieve reasonably accurate computations of the lift coefficient, Cheby-KAN exhibits considerably smaller errors than the MLP. Evidently, the model remains sufficiently accurate in computing integral quantities, thereby laying a solid foundation for its engineering applications.

\begin{figure} 
    \centering
    \includegraphics[width=14cm]{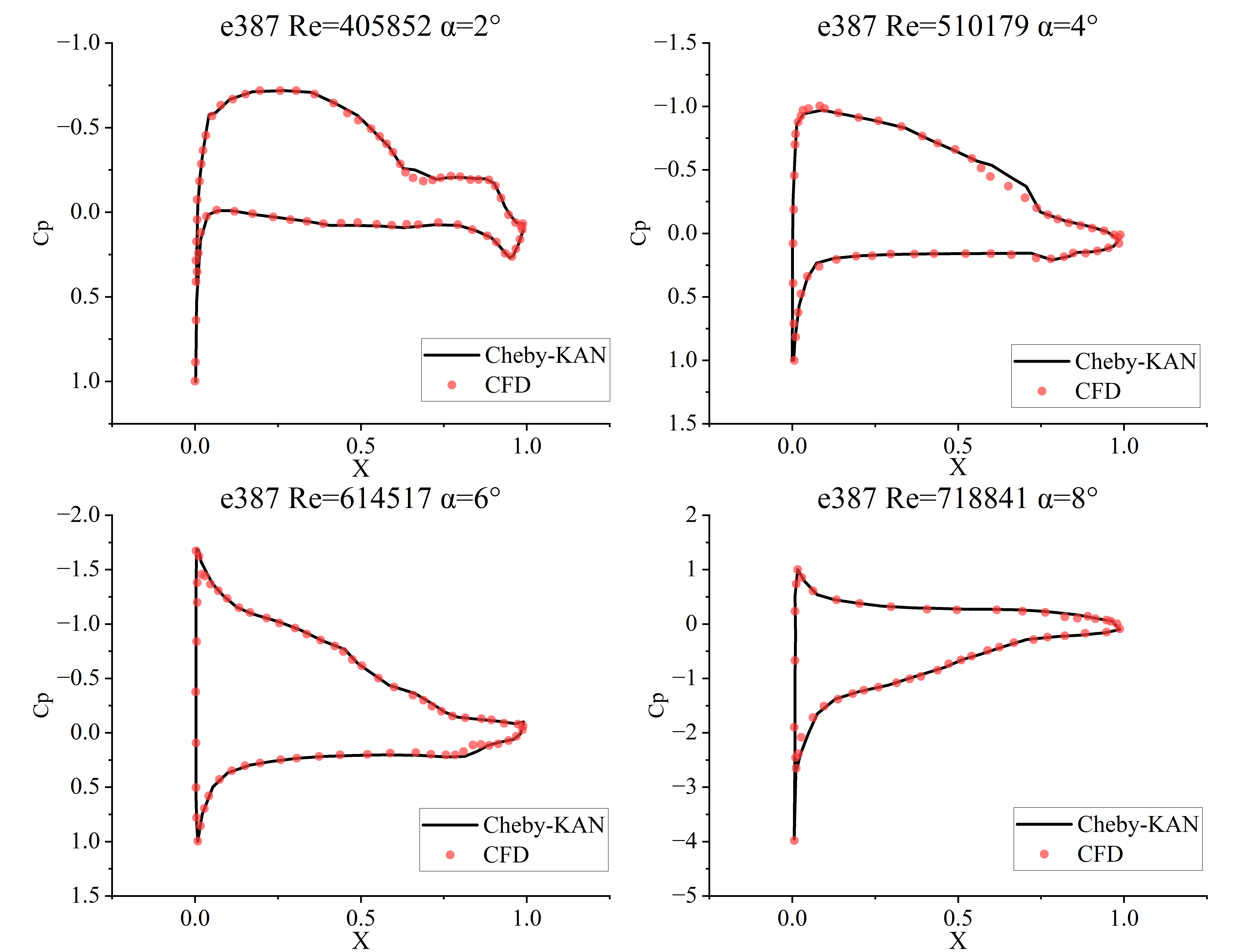}
    \caption{Comparison of CFD calculations and Cheby-KAN predicted CP curves for several operating conditions}
    \label{fig:Fig16}
\end{figure}

To evaluate the model's fitting performance and practical applicability on the upper and lower surfaces of the airfoil, Figure \ref{fig:Fig16} compares the pressure coefficient ($C_p$) curves obtained from CFD simulations with those predicted by the Cheby-KAN network under four representative operating conditions, using normalized pressure data. The results indicate that the Cheby-KAN predictions align closely with the CFD results over most of the airfoil surface. However, discrepancies are observed at isolated points, with the magnitude and frequency of errors varying across conditions. The mean absolute error across all test cases is approximately 10\%, which is consistent with the $R^2$ evaluation.

Meanwhile, the $C_p$ curve reveals a notable limitation of the current model: while it effectively captures the global trends of the dataset, its predictive accuracy diminishes in regions characterized by small disturbances or localized flow variations. This is particularly evident in portions of the $C_p$ distribution where steep pressure gradients are present, where the model tends to exhibit higher prediction errors. Improving the model’s performance in such sensitive regions will constitute a key focus of future work.

\begin{figure} 
    \centering
    \includegraphics[width=14cm]{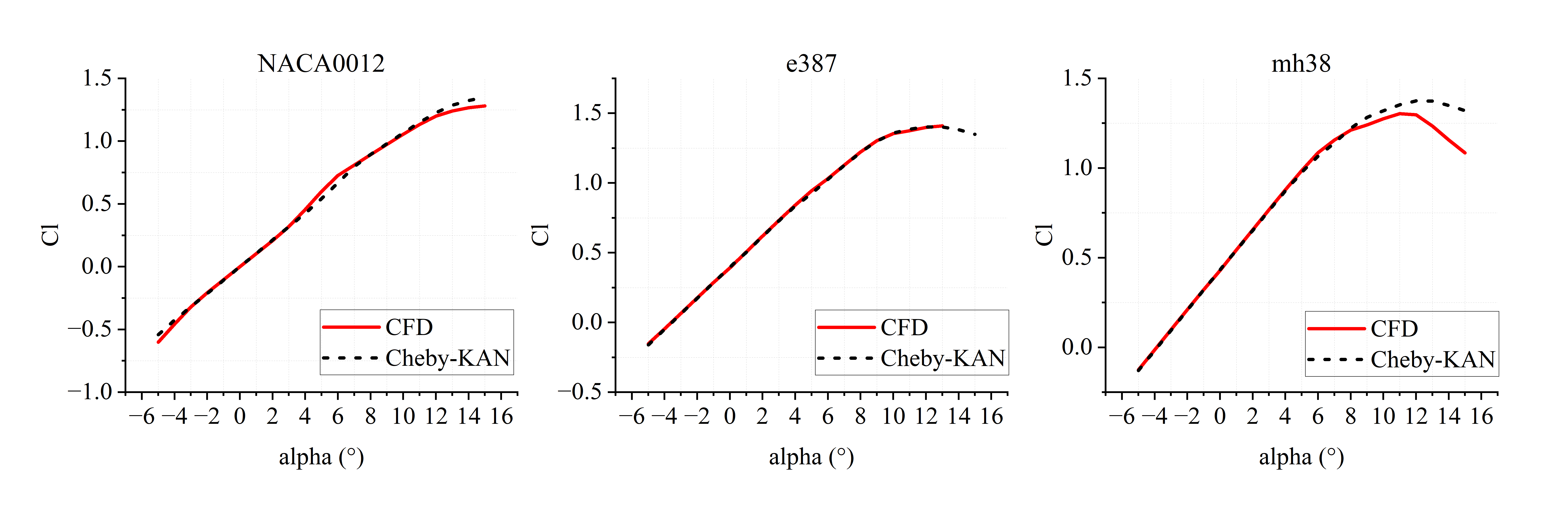}
    \caption{Lift characteristic diagrams of three selected airfoils}
    \label{fig:Fig17}
\end{figure}

Figure \ref{fig:Fig17} presents the lift characteristics of three selected airfoils at various angles of attack. The results indicate that the model accurately predicts the lift coefficient within the $-5^\circ - 10^\circ$ angle of attack range. However, beyond this range, the model struggles to capture the decrease in lift coefficient caused by flow separation—a limitation attributable to the absence of embedded physical mechanisms and the underrepresentation of separated flow conditions in the training dataset. Nevertheless, within the training regime, the model predicts lift coefficients with good accuracy and exhibits some capacity for mild extrapolation.

\begin{figure} 
    \centering
    \includegraphics[width=8cm]{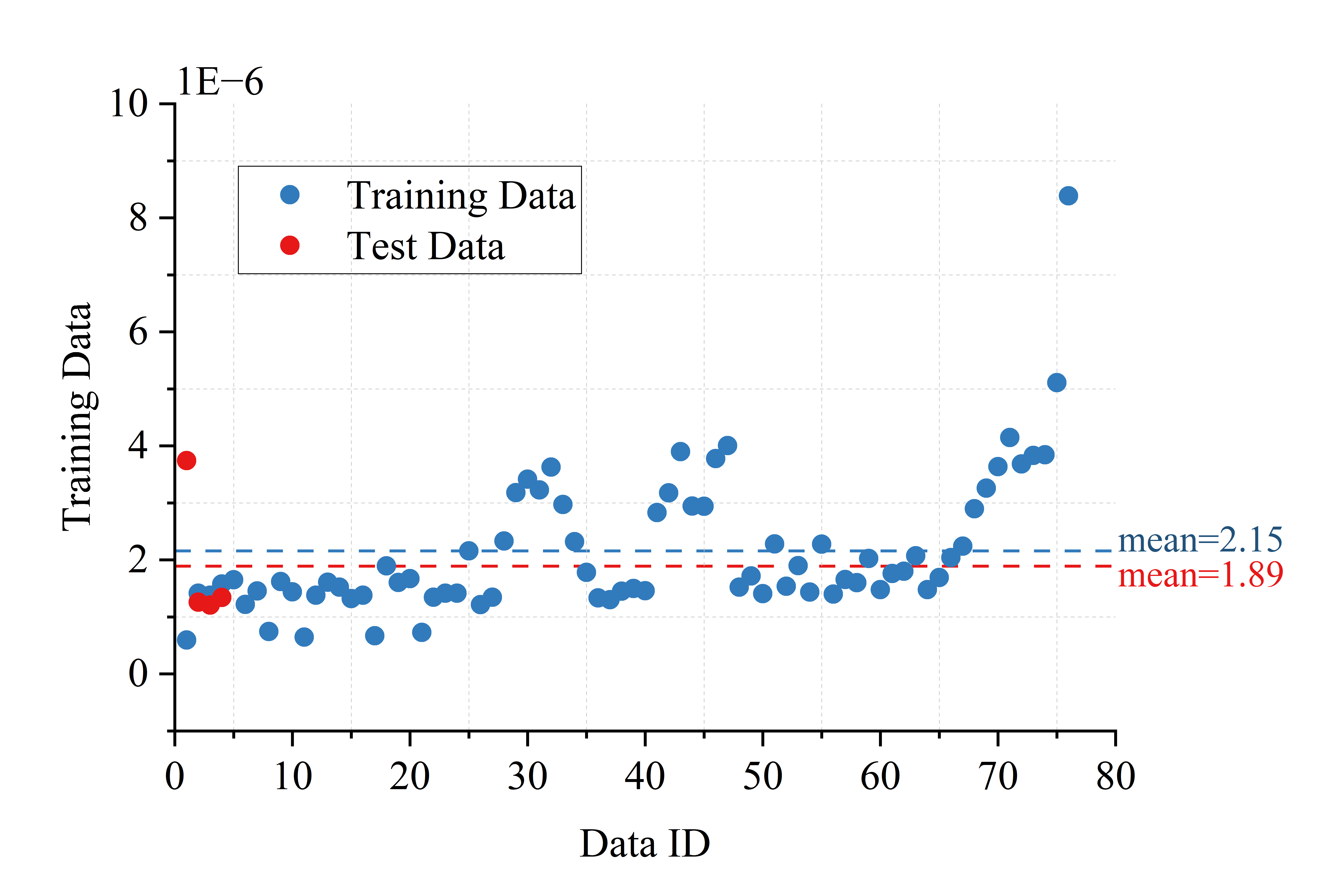}
    \caption{Comparison of MSE distribution between training and test sets}
    \label{fig:Fig18}
\end{figure}

To evaluate the model's generalization capability, assess its robustness, and examine potential overfitting, the model was tested on multiple flow field datasets that were excluded from the training process. The mean squared error (MSE) values from these unseen datasets were compared with those from the training data, yielding two distributions of MSE values, as shown in Figure \ref{fig:Fig18}.

The results demonstrate a negligible difference between the MSE values of the training and test sets. The calculated average MSE is $2.15 \times 10^{-6}$ for the training set and $1.89 \times 10^{-6}$ for the test set, confirming the model's strong generalization capability with no indication of overfitting.

This comprehensive analysis demonstrates the effectiveness and robustness of the proposed hybrid CNN-Cheby-KAN architecture for predicting airfoil pressure distributions. The model achieves high predictive accuracy across various airfoil types and operating conditions while maintaining strong generalization capabilities without evidence of overfitting. Our systematic investigation reveals that a dataset size of 300,000 samples represents an optimal balance between computational efficiency and model performance, beyond which marginal returns diminish significantly. The comparative analysis confirms Cheby-KAN's superior performance over traditional MLP architectures in capturing complex aerodynamic relationships, particularly in computing integral quantities like lift coefficients with remarkable precision. Although localized errors persist in regions of high curvature, specifically at the leading edge, the overall model performance remains satisfactory for engineering applications. These findings establish a solid foundation for future work aimed at addressing current limitations while highlighting the potential of Cheby-KAN networks as powerful tools in computational fluid dynamics and aerodynamic design optimization.

\section{Conclusion}
This study has presented a comprehensive evaluation of a novel hybrid CNN-Cheby-KAN framework for the rapid and accurate prediction of two-dimensional airfoil pressure distributions. Through rigorous experimentation and comparative analysis against traditional MLPs, the following key conclusions are drawn:
\begin{enumerate}
  \item A hybrid CNN-Cheby-KAN deep learning framework is successfully developed, demonstrating high accuracy in predicting two-dimensional airfoil pressure distributions, achieving a mean squared error on the order of $10^{-6}$ and a coefficient of determination ($R^2$) exceeding 0.999.
  \item The proposed model significantly outperforms traditional Multilayer Perceptrons (MLPs) in predictive accuracy and generalizability across various airfoils and operating conditions, particularly in computing integral aerodynamic quantities like lift coefficients with remarkable precision.
  \item Systematic investigation reveals that a dataset size of 300,000 samples represents the optimal balance between computational expense and model performance for the Cheby-KAN architecture, beyond which marginal accuracy gains diminish significantly.
  \item The model exhibits excellent generalization capability with no evidence of overfitting, as confirmed by nearly identical error distributions between training and test sets.
  \item While the model demonstrates acceptable overall accuracy for engineering applications, it exhibits certain limitations in localized error control and extrapolation capability, particularly under flow separation conditions. These shortcomings primarily stem from the absence of embedded physical constraints and the limited number of flow separation cases included in the training dataset.
\end{enumerate}

\bibliographystyle{ieeetr}  
\bibliography{references}

\end{document}